\documentclass[%
 reprint,
 amsmath,amssymb,
pra,
]{revtex4-1}
\usepackage[utf8]{inputenc}
\usepackage[left=1in,right=1in,top=1in,bottom=1in]{geometry}
\usepackage{amsmath}
\usepackage{amsthm}
\usepackage{amsfonts}
\usepackage{amssymb}
\usepackage{mathtools}
\usepackage{algorithm}
\usepackage[noend]{algpseudocode}
\usepackage{float}
\usepackage{verbatim}
\usepackage{ctable}

\usepackage{lineno}
\usepackage{graphicx}
\usepackage{dcolumn}
\usepackage{bm}

\begin{document}

\title{Understanding cytoskeletal avalanches using mechanical stability analysis}

\author{Carlos Floyd}
\affiliation{%
	Biophysics Program, University of Maryland, College Park, MD 20742 USA
}%

\author{Herbert Levine}
\affiliation{
	Department of Bioengineering}
\affiliation{Department of Physics, Northeastern University, Boston, MA 02115 USA}

\author{Christopher Jarzynski}
\email{cjarzyns@umd.edu}
\affiliation{
	Department of Chemistry and Biochemistry,}
\affiliation{
	Institute for Physical Science and Technology,}
\affiliation{Department of Physics, University of Maryland, College Park, MD 20742 USA}

\author{Garegin A. Papoian}
\email{gpapoian@umd.edu}
\affiliation{Department of Chemistry and Biochemistry}
\affiliation{
	Institute for Physical Science and Technology, University of Maryland, College Park, MD 20742 USA}

\date{\today}

\begin{abstract}

Eukaryotic cells are mechanically supported by a polymer network called the cytoskeleton, which consumes chemical energy to dynamically remodel its structure.  Recent experiments \textit{in vivo} have revealed that this remodeling occasionally happens through anomalously large displacements, reminiscent of earthquakes or avalanches.  These cytoskeletal avalanches might indicate that the cytoskeleton's structural response to a changing cellular environment is highly sensitive, and they are therefore of significant biological interest.  However, the physics underlying ``cytoquakes'' is poorly understood.  Here, we use agent-based simulations of cytoskeletal self-organization to study fluctuations in the network's mechanical energy.  We robustly observe non-Gaussian statistics and asymmetrically large rates of energy release compared to accumulation in a minimal cytoskeletal model.  The large events of energy release are found to correlate with large, collective displacements of the cytoskeletal filaments.  We also find that the changes in the localization of tension and the projections of the network motion onto the vibrational normal modes are asymmetrically distributed for energy release and accumulation.  These results imply an avalanche-like process of slow energy storage punctuated by fast, large events of energy release involving a collective network rearrangement.  We further show that mechanical instability precedes cytoquake occurrence through a machine learning model that dynamically forecasts cytoquakes using the vibrational spectrum as input.  Our results provide the first connection between the cytoquake phenomenon and the network's mechanical energy and can help guide future investigations of the cytoskeleton's structural susceptibility.

\end{abstract}

\pacs{Valid PACS appear here}
\maketitle


\section*{Introduction}

The actin-based cytoskeleton is an active biopolymer network that plays a central role in cell biology, providing the cell with a means to control its shape and produce mechanical forces during processes such as migration and cytokinesis \cite{fletcher2010cell, boal2012mechanics, mogilner1996cell, mogilner2006edge,wang2006introductory}.  These cellular-level forces arise from the collective non-equilibrium activity of molecular motors interacting with the actin filament scaffold, enabling dynamic, driven-dissipative cytoskeletal remodeling \cite{tyleramclaughlin2016collective, toyota2011non, mackintosh2008nonequilibrium}.  Recent experimental efforts have uncovered a remarkable phenomenon exhibited by cytoskeletal networks \textit{in vivo}: these networks undergo large, sudden structural rearrangements significantly more frequently than predicted by a Gaussian distribution \cite{alencar2016non, shi2019dissecting}.  Heavy-tailed distributions of event sizes are well-known in seismology, where the Gutenberg-Richter law describes the power-law relationship between the energy released by an earthquake and such an earthquake's frequency \cite{gutenberg1949seismicity, bak2002unified}.  Due to this analogy the term ``cytoquake,'' which we adopt here, has been coined by experimenters to describe large cytoskeletal remodeling events.  In previous work we have reported the first \textit{in silico} observations of this phenomenon, appearing as heavy tails in the distributions of mechanical energy released by cytoskeletal networks \cite{floyd2019quantifying}.  These findings suggest that avalanche-like processes may play a fundamental role in cytoskeletal dynamics. 

The physics underlying cytoquakes is not well understood, as current explanations based on experimental data are mostly speculative and rely on qualitative comparisons to systems amenable to computational study which similarly exhibit non-exponential relaxation, such as jammed granular packings and spin glasses \cite{alencar2016non, shi2019dissecting, van2009jamming, bouchaud1992weak}.  In particular, it is not known whether large cytoskeletal displacements actually arise from an avalanche-like process of slow energy storage and fast, large events of energy release.  Alternative explanations of heavy-tailed distributions of cytoskeletal displacements that do not involve avalanche-like dynamics have also been considered.  For instance, heterogeneity in the spatial distribution of molecular motors has been proposed as a possible mechanism for non-Gaussian distributions of displacements \cite{toyota2011non}.  Here, we describe the first detailed numerical study focused on the mechanical energy of cytoskeletal networks exhibiting large displacements.  We find that the statistics of energy accumulation and release support the hypothesis of avalanche-like dynamics occurring in cytoskeletal networks, and thus that avalanche-like dynamics do at least contribute to the observed heavy-tailed distributions of cytoskeletal displacements. 

In addition, in previous studies little emphasis has been given to the possible biological roles played by cytoquakes.  We propose one such role, that these large mechanical fluctuations are concomitant with a large susceptibility to mechanical forces or chemical perturbations, allowing the cytoskeleton to be highly sensitive to physiological cues arriving via various cell signaling pathways \cite{zhuravlev2009molecular}.  Dynamic instability is already an acknowledged feature of certain cytoskeletal components such as microtubules and filopodia \cite{mitchison1984dynamic}.  A similar design principle may also apply to larger cytoskeletal structures to allow fast remodeling.  For instance, avalanche-like dynamics may serve a useful purpose in the lamellipodia of migrating cells, which probe local chemical gradients and must quickly collapse protrusions in unsuccessful search directions as well as adaptively remodel their structure in response to changing mechanical loads \cite{boal2012mechanics,mueller2017load}.   However, to investigate such possible biological roles we first need a more detailed account of the underlying causes of the observed large structural rearrangements, which is the subject of this paper.  

Here, we perform detailed simulations of a minimal cytoskeletal model system using the software package MEDYAN (Mechanochemical Dynamics of Active Networks) \cite{popov2016medyan}.  Our main qualitative result is that there is a significant asymmetry between how cytoskeletal networks accumulate and release mechanical energy.  While both accumulation and release statistics are heavy-tailed, the magnitudes of energy release are more broadly distributed than those of energy accumulation.  Several measures of network dynamics are also found to be distributed asymmetrically for energy release and accumulation, including the network displacement, the localization of tension, and the projection of the network motion onto the vibrational normal modes.  These results support an avalanche-like picture of slow energy accumulation punctuated by fast, broadly-distributed events of energy release that involve a collective structural rearrangement of the network.  The asymmetric energy fluctuations are found to be robust against changes in chemical concentrations and system size, suggesting that avalanches are intrinsic to cytoskeletal network dynamics.  We further establish a connection between cytoquakes and mechanical stability, both through the observed spatial delocalization of tension during cytoquakes and the machine learning-assisted ability to dynamically forecast cytoquakes using the Hessian eigenspectrum of the mechanical energy function.  This implies that mechanical instability, as encoded in the Hessian eigenspectrum, precedes incipient cytoquakes which then act to homogenize tension in the network.  At the end of the paper we pose several open questions based on these results, which can help to guide future investigations into cytoquakes and their possible physiological functions.

\section*{Results}
\subsection*{Energy fluctuations are asymmetric, heavy-tailed, and self-affine}

\begin{figure}[ht]
	\centering
	\includegraphics[width=1.0\linewidth]{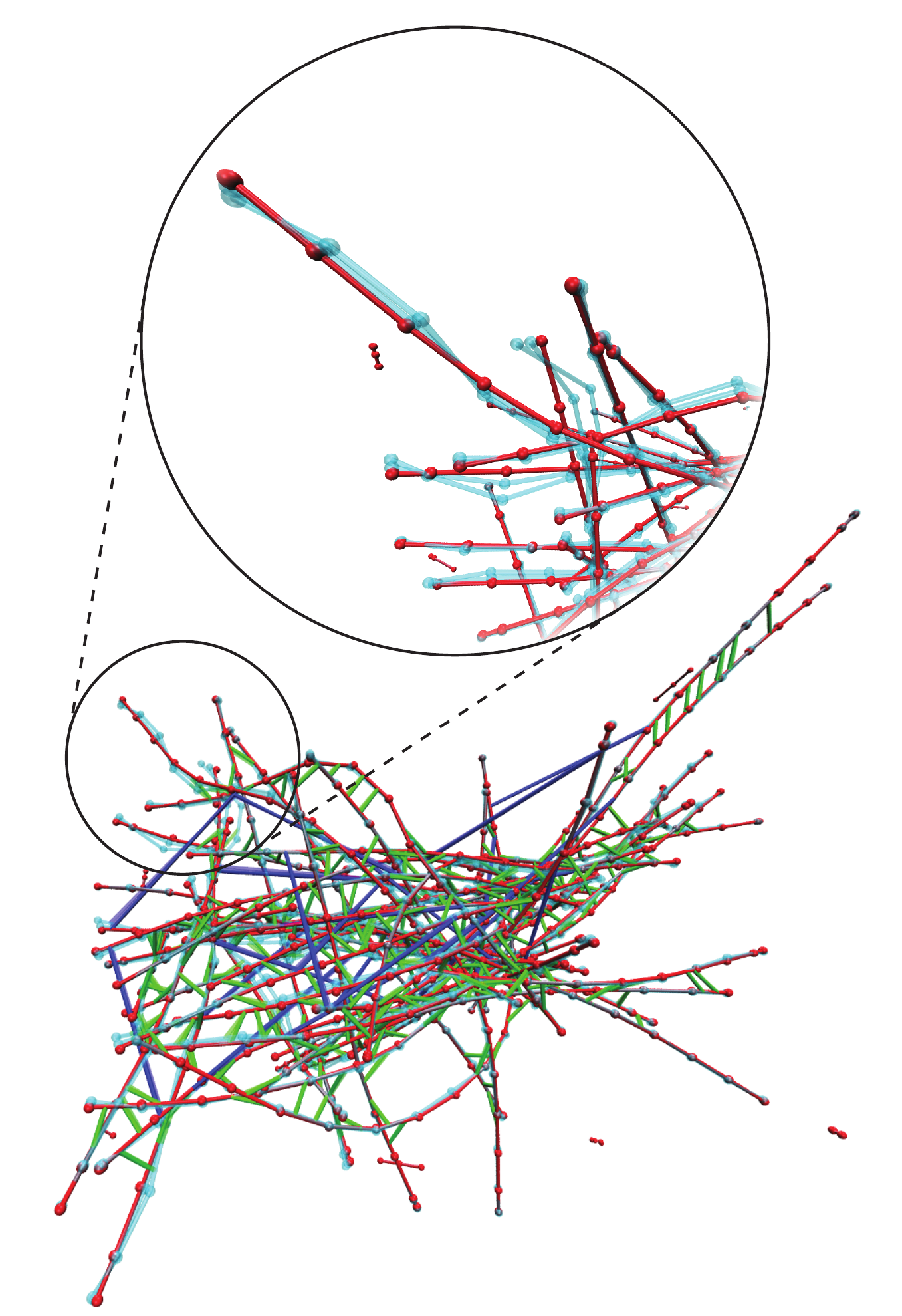}
	\caption{A snapshot from a MEDYAN trajectory of an actomyosin network in a $1 \ \mu m^3$ box for the condition $C_{3,3}$ (see Materials and Methods).  Actin filaments are shown in red, $\alpha$-actinin is shown in green, and myosin motors are shown in blue.  Beads representing the joined points (i.e. hinges) of thin cylinders (at most $54 \ nm$ long) are visualized as red spheres.  The cyan filaments represent motion of the network corresponding to a soft, delocalized vibrational mode determined from Hessian analysis, as described in the main text.  In the inset we zoom in on part of the network and exclude associated proteins to show greater detail of this vibrational motion. }
	\label{network}
\end{figure}

We study a subsystem of the full cytoskeleton called an actomyosin network.  This consists of semi-flexible actin filaments and associated proteins, including active molecular motors (e.g. minifilaments of non-muscle myosin IIA) and passive cross-linkers (e.g. $\alpha$-actinin).  An actomyosin network as represented in simulation is visualized in Figure \ref{network}.  The actin filaments hydrolyze ATP molecules in a directed polymerization process which reaches a steady state called ``treadmilling'' \cite{floyd2017low}.  The myosin minifilaments ($\sim$200 $nm$ in length) transiently bind to pairs of actin filaments and also hydrolyze ATP as fuel to walk along the filaments, generating motion and mechanical stresses.  These active process drive the network away from equilibrium.  The cross-linkers ($\sim$35 $nm$) bind more stably to nearby filaments, serving to transmit the force produced by motors and to both store and through unbinding dissipate the resulting energy, heating the environment \cite{kovacs2003functional, erdmann2013stochastic, howard2001mechanics,  otey2004alpha,  komianos2018stochastic, lieleg2008transient, kurzawa2017dissipation}.  Dissipation of stored mechanical energy also occurs as filaments relax out of strained configurations, in a manner which depends on mutual constraints filaments exert on each other through bound cross-linkers and motors.  Additionally, the rates of motor walking and unbinding as well as of cross-linker unbinding depend exponentially on the forces sustained by these molecules, giving rise to nonlinear coupling between the mechanical state of the network and its chemical propensities \cite{keller2000mechanochemistry, pereverzev2005two}.  These processes by which the ability of the network to mechanically relax depends on its current state set the stage for avalanche-like dynamics.  

Using MEDYAN, we performed simulations of small cytoskeletal networks consisting of 50 actin filaments in 1 $\mu m^3$ hard-walled cubic boxes with varying concentrations of $\alpha$-actinin cross-linkers ($[\alpha]$) and of NMIIA myosin motor minifilaments ($[M]$)  \cite{popov2016medyan, floyd2019quantifying, chandrasekaran2019remarkable, ni2019turnover, li2020tensile}.  We omit here other associated proteins, such as the branching agent Arp2/3, finding that our minimal system is sufficient to produce heavy-tailed distributions of event sizes, although it has recently been discovered that branching acts to enhance avalanche-like processes \cite{liman2020role}.  MEDYAN simulations combine stochastic chemical dynamics with a mechanical representation of filaments and associated proteins (see the SI Appendix, Description of MEDYAN simulation platform for a detailed outline of the MEDYAN model).  Simulations proceed iteratively in a cycle of four steps: 1) stochastic chemical simulation for a time $\delta t$ (here $0.05 \ s$), 2) computation of the resulting new forces, 3) equilibration via minimization of the mechanical energy, and 4) updating of force-sensitive reaction rates such the as slip-bonds of cross-linkers, catch-bonds of motors, and motor stalling.  Recent extensions to the MEDYAN platform allow calculation of the change in the system's Gibbs free energy during each of these steps \cite{floyd2019quantifying, floyd2020gibbs}, originally applied to study the thermodynamic efficiency of myosin motors in converting chemical free energy to mechanical energy under various conditions of cross-linker and motor concentration.  We employ this methodology here and focus on the statistics of the system's mechanical energy $U$ as it self-organizes.   

\begin{figure}[ht]
	\centering
	\includegraphics[width=1.0\linewidth]{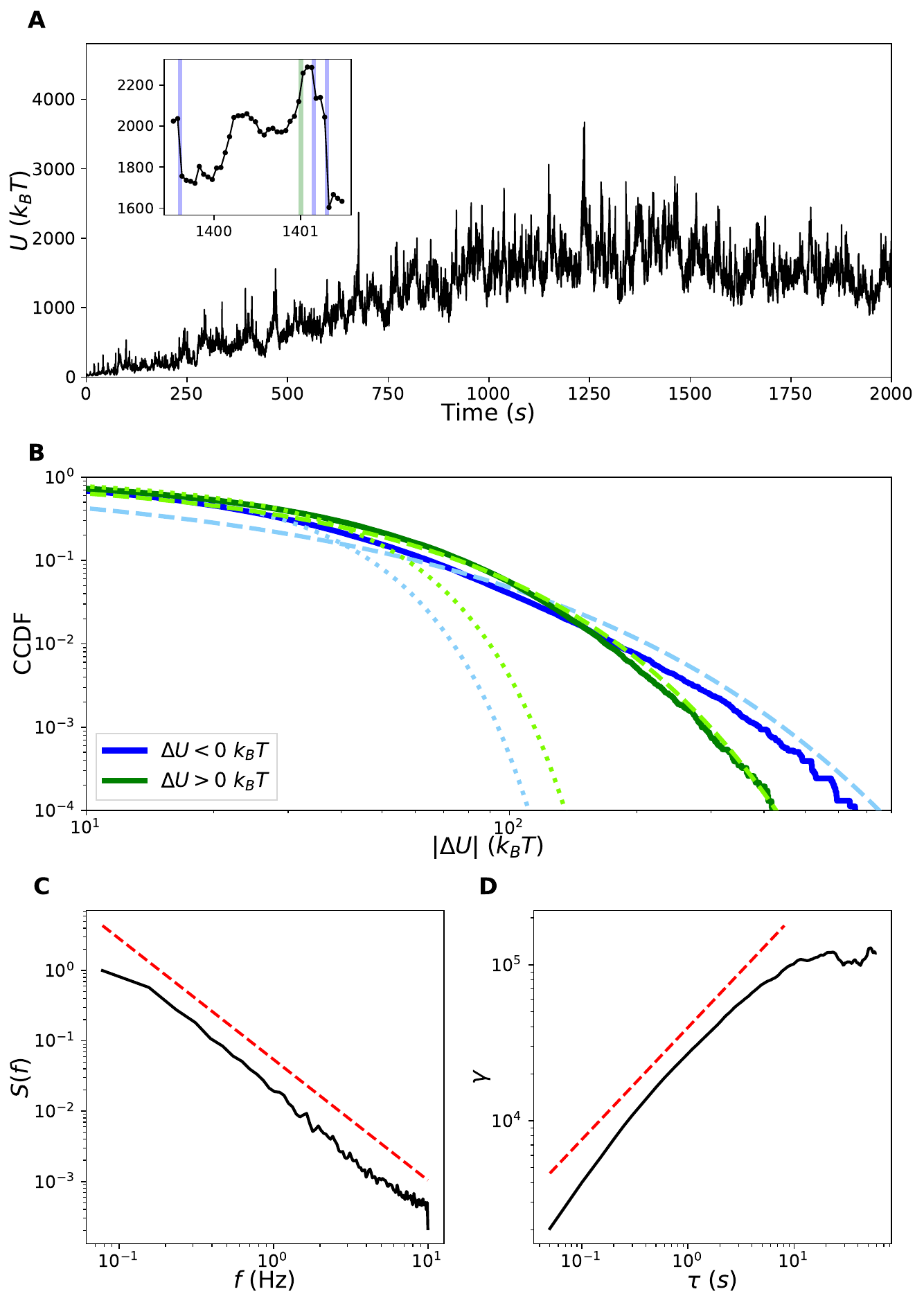}
	\caption{Statistics of $\Delta U$. \textbf{A}: Trajectory of the network's mechanical energy $U(t)$ for condition $C_{3,3}$ (see Materials and Methods).  Inset: A blow-up of the trajectory to show instances of rare events of energy release ($\Delta U < - 100 \ k_B T$, blue) and accumulation ($\Delta U > 100 \ k_B T$, green).  \textbf{B}:  CCDFs of $|\Delta U_-|$ (blue) and $\Delta U_+$ (green) collected from five runs when the system is at QSS after 1000 $s$.  Dotted lines in lighter colors represent fits to the data of a half-normal CCDF, and dashed lines represent fits of stretched exponentials. \textbf{C}: The normalized power spectral density of $U(t)$ for a single run at QSS from which the spectral exponent $\beta = 1.72$ is determined by fitting a power-law, shown offset in red.  \textbf{D}: The semivariogram   obeys the scaling relationship $\gamma \sim \tau^{2H_a}$ over the scaling range.  }
	\label{TrajCCDF}
\end{figure}

We first characterize the observed occurrence of avalanche-like dynamics in these simulations.  The simulations begin with short seed filaments that quickly polymerize (tens of seconds) to their steady-state lengths. Following this, the slower process (hundreds of seconds) of primarily myosin-driven self-organization occurs which for most conditions results in geometric contraction to a percolated network (see Movie 1) \cite{ komianos2018stochastic, wang2012active}.  The mechanical energy $U(t)$ fluctuates near a quasi-steady state (QSS) value, which we analyze as a stochastic process. In Figure \ref{TrajCCDF}.A we display the trajectory of $U(t)$ for condition $C_{3,3}$ (with $\alpha$-actinin concentration $[\alpha] = 2.82$ $\mu M$, and motor concentration $[M] = 0.04$ $\mu M$; see the Materials and Methods for a description of the experimental conditions).  We tracked the net changes of the mechanical energy $\Delta U(t) = U(t + \delta t) - U(t)$ resulting from each complete cycle of simulation steps 1) - 4).  For the purpose of analyzing the observed asymmetric heavy tails in the distribution of $\Delta U$, we treat the negative increments $\Delta U_-$ (energy release) and positive increments $\Delta U_+$ (energy accumulation) as samples from separate distributions with semi-infinite domains.  The complementary cumulative distribution functions (CCDFs or ``tail distribution'', the probability $P(X \geq x)$ of observing a value of the random variable $X$ above a threshold $x$, as a function of $x$) of the observed samples collected from all five runs at QSS are illustrated in Figure \ref{TrajCCDF}.B.  Both distributions display striking heavy tails relative to a fitted half-normal distribution.  The CCDFs are better fit by stretched exponential (Weibull) functions of the form \cite{murthy2004weibull}
\begin{equation}
P(X \geq x) =  e^{-(x/\lambda)^k}. 
\end{equation}  
We justify this choice of distribution by constructing Weibull plots, as discussed in the SI Appendix, Weibull plots.  We find $k = 0.60 \pm 0.06$ for $|\Delta U_-|$ and $k = 0.83 \pm 0.07$ for $\Delta U_+$ with uncertainty taken over the five runs, indicating shallower tails for energy release compared to energy accumulation. We also measured parameter $\eta$ that indicates non-Gaussianity:
\begin{equation}
\eta = \frac{\langle x^4 \rangle}{3 \langle x^2 \rangle ^2} - 1,
\label{eqngp}
\end{equation}
where $\langle x^m \rangle$ is the $m^\text{th}$ moment about zero; for a half-normal distribution $\eta = 0$, and $\eta > 0$ quantifies heavy-tailedness.  We find $\eta = 11.37 \pm 5.37$ for $|\Delta U_-|$ and $\eta =  1.96 \pm 0.58$ for $\Delta U_+$.  This, along with the shallower tails of the fitted stretched exponential functions, indicates greater deviation from Gaussianity for energy release compared to energy accumulation.  These results support the picture that typically energy accumulates comparatively slowly and is released via large occasional events. 

We next analyze the temporal correlations of $U(t)$ at QSS.  A self-affine stochastic time series $G(t)$, for which $G(t)$ and $|\zeta|^{H_a} G(t/\zeta)$ have the same statistics for any scaling parameter $\zeta$, has a power spectral density $S(f)$ exhibiting a power-law dependence on frequency $f$:
$
S(f) \propto f^{-\beta},
\label{eqFT}
$
where the spectral exponent $\beta$ is the persistence strength, related to the color of the signal \cite{malamud1999self, pelletier1999self}.  We find $\beta = 1.72 \pm 0.02$ for $U(t)$, as shown in Figure \ref{TrajCCDF}.C.  With this value of $\beta$, $U(t)$ is classified as a pinkish-brown signal, implying it is non-stationary and has temporally anti-correlated increments $\Delta U$.  Self-affine time series further obey the theoretical relationship $\beta = 2 H_a +1$ when $1 \leq \beta \leq 3$, where $H_a$ is the Hausdorff exponent determined from the scaling of the semivariogram 
\begin{equation}
\gamma(\tau) = \frac{1}{2}\overline{\left(G(t+\tau) - G(t)\right)^2}  \sim \tau^{2H_a},
\label{eqsemivariogram}
\end{equation}
and where the overbar represents temporal averaging \cite{turcotte1997fractals, hergarten2002self}.  We find that this relationship is satisfied by $U(t)$, as shown in Figure \ref{TrajCCDF}.D, yielding $H_a = 0.36 \pm 0.01$ and confirming that $U(t)$ is self-affine.  Such non-Markovian and self-affine time series and spatial patterns commonly arise in various complex geophysical processes (e.g. the temporal variation of river bed elevation), further supporting the analogy between the cytoskeleton and earth systems \cite{witt2013quantification, williams2019self}.

\subsection*{Distinguishing features of cytoquakes}

\begin{figure}[ht]
	\centering
	\includegraphics[width=1.0\linewidth]{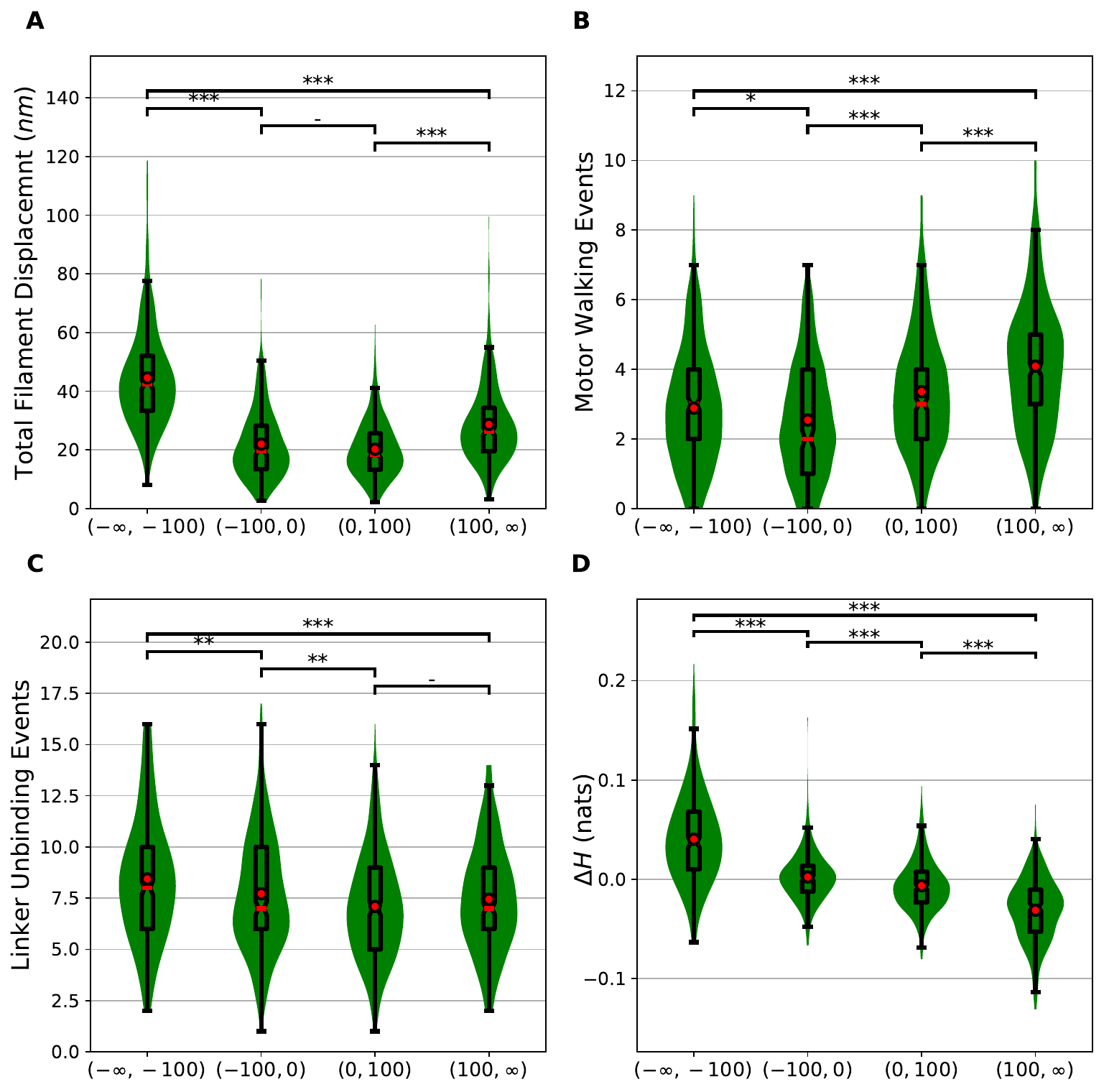}
	\caption{\textbf{A}: Differences in the total filament displacement between simulation cycles for which $\Delta U < -100 \ k_B T$, cycles for which $\Delta U \in (-100 \ k_B T, \ 0 \ k_B T)$, cycles for which $\Delta U \in ( 0 \ k_B T, \ 100 \ k_B T)$, and finally cycles for which $\Delta U > 100 \ k_B T$. To compare these distributions, the two-sided $p$-value of the Wilcoxon rank-sum test between pairs of cycle types is reported as being either not significant: - ($p\geq 0.05$), significant at level 1: * ($p<0.05$), at level 2: ** ($p<0.01$), or at level 3: *** ($p<0.001$) \cite{hogg2005introduction}.  Since there many more simulation cycles for the categories $\Delta U \in (-100 \ k_B T, 0 \ k_B T)$ and $\Delta U \in (0 \ k_B T, 100 \ k_B T)$, we took a random sub-sample ($\sim 300$ each) of all events for these categories to be roughly equal to the number of events for which $\Delta U < -100 \ k_B T$ and for which $\Delta U > 100 \ k_B T$. In these combination violin and box-and-whisker plots, the red circle represents the mean, the red bar represents the median, and the notches in the box represent the $95 \% $ confidence interval of the median.  \textbf{B}: Differences in the number of motor walking events between the different cycle types as just described.  \textbf{C}: Differences in the number of $\alpha$-actinin unbinding events between the different cycle types. \textbf{D}: Differences in the changes in Shannon entropy $\Delta H$ of the spatial tension distribution of network tension between the different cycle types.
	}
	\label{ViolinPlots}
\end{figure}

\begin{figure}[ht]
	\centering
	\includegraphics[width=1.0\linewidth]{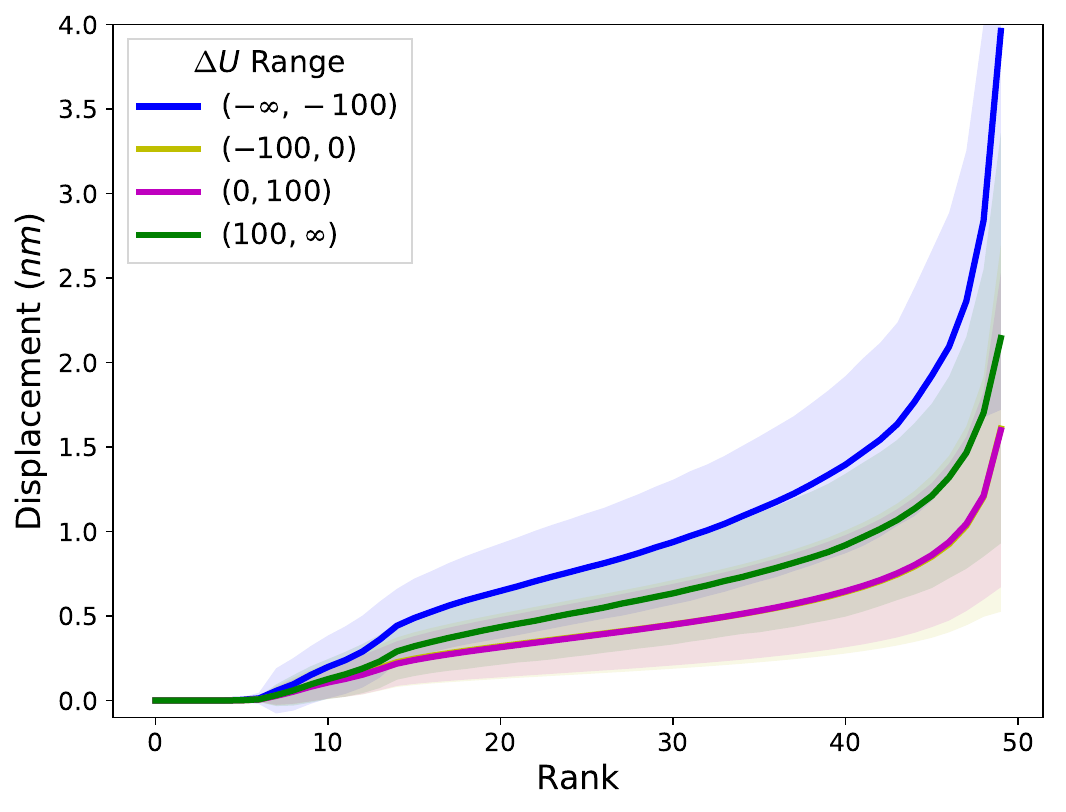}
	\caption{Rank-size distribution of the displacements experienced by each of the 50 filaments during simulation cycles when $\Delta U$ is in different ranges, in units of $k_B T$.  For each cycle, the filaments are ranked according to their displacement and these ranks are plotted against the corresponding displacement.  The average and standard deviation of these rank-displacement curves are taken over each cycle in a given category.  The curves for the categories $\Delta U \in (-100 \ k_B T, \ 0 \ k_B T)$ and $\Delta U \in ( 0 \ k_B T, \ 100 \ k_B T)$ are nearly coincident.  This data is collected from one run of condition $C_{3,3}$ at QSS.}
	\label{RankDist}
\end{figure}

We find that cytoquakes, defined throughout as simulation cycles for which $\Delta U < -100 \ k_b T$ (chosen to lie well in the tail of the distribution of $\Delta U$, see Figure \ref{TrajCCDF}), are correlated with several changes in the state of the network.  In Figure \ref{ViolinPlots} we show that rare large events of energy accumulation correspond to a greater than usual number of myosin motor steps whereas rare large events of energy release correspond to greater than usual total displacement of the actin filaments and a slightly greater number of linker unbinding events.  The displacement between filaments from $t$ to $t + \delta t$ is calculated by triangulating the area between the two filament configurations and dividing the area by the filament length, as described in SI Appendix, Filament displacements. The total filament displacement at time $t$ is computed as the sum of displacements over all filaments during the time interval $(t, t+\delta t)$. This quantity is found to be largest during cytoquake events.  Furthermore, these large total displacements do not come from highly localized motions.  Instead, they depend on many filaments each displacing an unusually large amount, as shown in Figure \ref{RankDist} where the filaments are ranked according to their displacement during a cycle.  For cytoquake events, the typical displacement at almost every rank is greater than the corresponding displacement at that rank for other cycle types.  This agrees with the notion of cytoquakes as a large and collective structural rearrangement of the network.  

We also observe cytoquakes to induce a spatial homogenization of the tension sustained by the network during large events of energy release, as quantified by changes in the Shannon entropy of the spatial tension distribution $H(t)$ (see Figure \ref{ViolinPlots}.D).  The tension distribution $P_{ijk}$ is constructed by discretizing the simulation volume of $1 \ \mu m^3$ into a grid of $10^3$ voxels indexed by $i,j,k$, and computing the proportion of the total network tension belonging to the mechanical elements (filament cylinders, cross-linkers, and motors) inside each voxel.  Additional details for the calculation of $H(t)$ can be found in Materials and Methods.  The combination of large, collective rearrangement and a spatial homogenization of tension supports the interpretation of cytoquakes as an avalanche-like event of energy release.

\subsection*{Asymmetric statistics are robust across concentrations and system-size}

We next discuss how these results generalize to different concentrations of associated proteins and different system sizes.  Five concentrations of $\alpha$-actinin (ranging from 0.17 to 5.48 $\mu M$) and five concentrations of myosin miniflaments (ranging from 0.003 to 0.08  $\mu M$) were tested with a constant G-actin monomer concentration of 13.3 $\mu M$, in the regime of physiological concentrations \cite{milo2015cell}.  At the lowest concentrations of cross-linkers and motors, the network did not contract, representing a very different actomyosin phase to which we omit comparisons.  For all of the conditions producing contracting networks, we found that asymmetric heavy-tailed distributions of $\Delta U$ persist, with large values of the non-Gaussian parameter for $|\Delta U_-|$ ($\eta \sim 5 - 20$) and $\Delta U_+$ ($\eta \sim 2 - 5$), although $\eta$ for negative increments was observed to decrease with the motor concentration (SI Appendix, Figure S1).  We conclude that the avalanche-like energy fluctuations discussed above are not highly sensitive to associated protein concentrations.  These fluctuations may depend on the parameters of the force-sensitive reaction rates (which are taken here to correspond to experimental values), but we leave this interesting question for future work.

\begin{figure}[ht]
	\centering
	\includegraphics[width=1.0\linewidth]{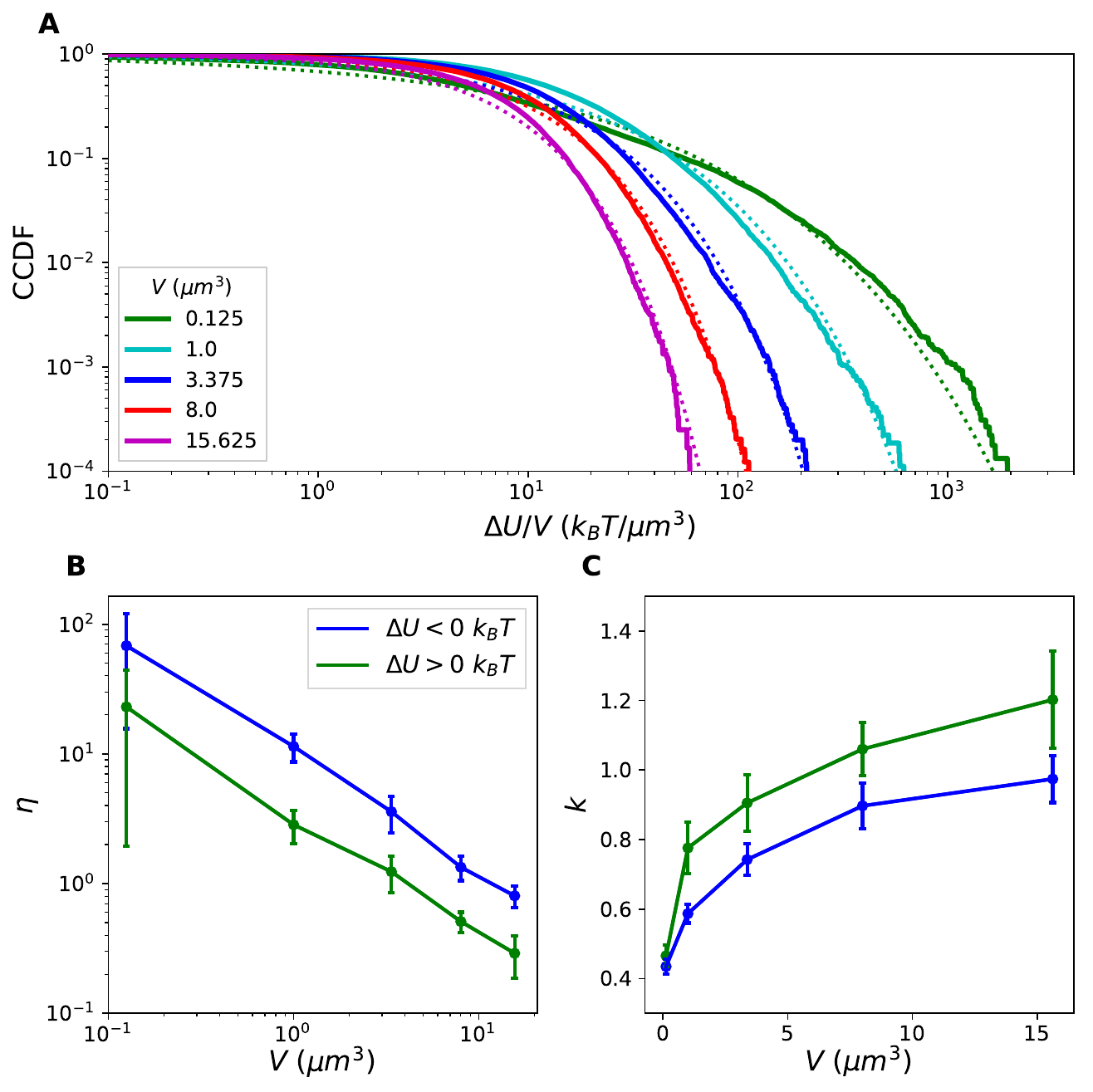}
	\caption{\textbf{A}: CCDFs of $|\Delta U_-|$ normalized by the system volume $V$ collected for 5 runs for increasing system sizes plotted against the fitted stretched exponential functions. \textbf{B}: The non-Gaussian parameter $\eta$ for $|\Delta U_-|$ and $\Delta U_+$ are plotted with uncertainty taken over the different runs.  \textbf{C}: The Kohlrausch exponents $k$ for $|\Delta U_-|$ and $\Delta U_+$.}
	\label{FS}
\end{figure}

We performed a finite-size scaling study by holding the concentrations of condition $C_{3,3}$ fixed (with $\alpha$-actinin concentration $[\alpha] = 2.82$ $\mu M$ and motor concentration $[M] = 0.04$ $\mu M$) and varying the system volume $V$.  Larger systems reach QSS at later times, and our simulations of larger systems did not reach QSS in the allotted computational time.  As a result, we collected samples of $\Delta U$ for these systems on the approach to QSS, from $300 $  to $800 \ s$, once the networks had all nearly fully percolated (i.e. nearly all filaments belonged to a single component connected by cross-linkers), trusting that the relevant scaling behavior could still be observed.  Stretched exponential functions approximately fit the distributions of $\Delta U_+$ and $|\Delta U_-|$ for all system sizes (see Figure \ref{FS}.A for the fits of $|\Delta U_-|$).  Larger systems displayed steeper tails as indicated by the observed power-law decay of $\eta$ for $|\Delta U_-|$ and $\Delta U_+$ (Figure \ref{FS}.B), although interestingly $\eta$ for $|\Delta U_-|$ is larger than that for $\Delta U_+$ by a constant factor of roughly 3 for all systems sizes.  The steeper tails are also evidenced by the slow growth of the Kohlrausch exponents $k$ with $V$ (Figure \ref{FS}.C).  Thus, the distributions of energy release and accumulation across the entire network become narrower and more Gaussian for large systems.  This, in contrast to driven-dissipative systems that exhibit self-organized criticality, suggests the existence of some intrinsic and finite scale for avalanche-like releases of energy in cytoskeletal networks.   By summing over many local energy fluctuations of this finite scale, the distribution of the fluctuations in the total energy $U$ becomes increasingly Gaussian for large systems owing to the central limit theorem.  This intrinsic scale may be partly determined by the non-conservative transfer (dissipation) of mechanical energy as it spreads through the network during avalanches \cite{howard2001mechanics,pun2020prediction}. 

\subsection*{Local vs. global metrics}

\begin{figure}[ht]
	\centering
	\includegraphics[width=1.0\linewidth]{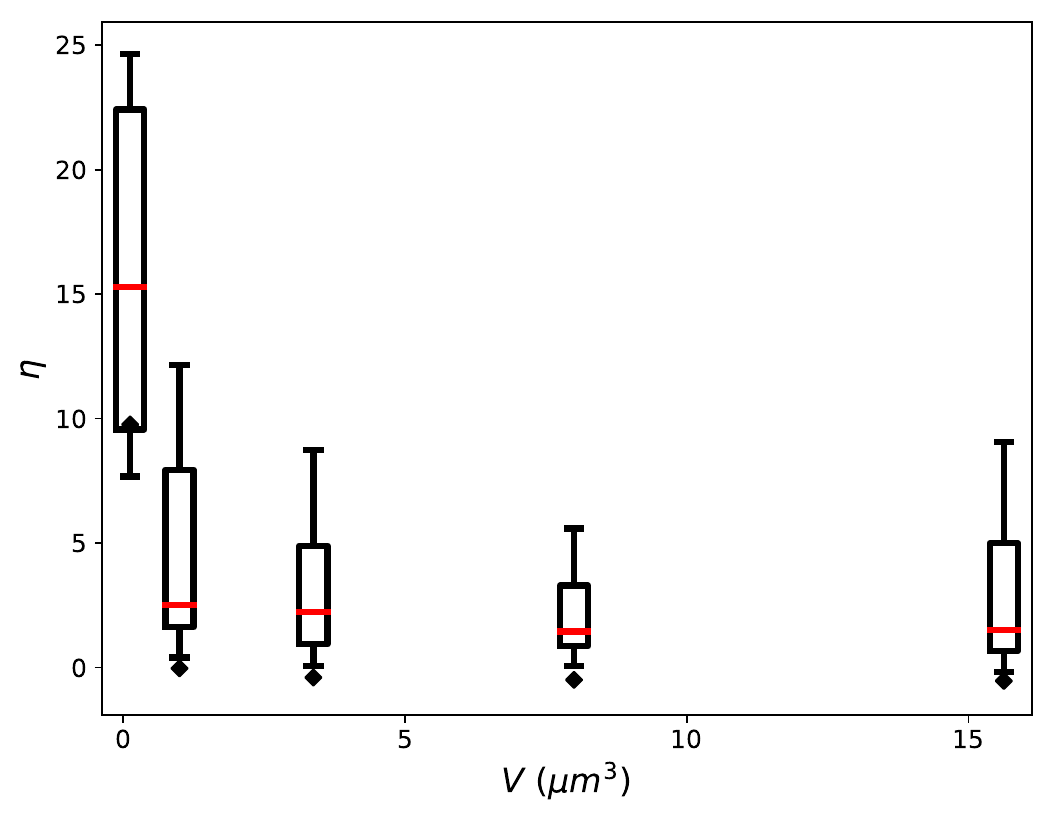}
	\caption{Plots for different simulation volumes $V$ of the distributions of non-Gaussian parameter $\eta_f$ of the distributions of individual filament displacements.  The box and whisker plots summarize the distribution of $\eta_f$ for all filaments in the system, with the median shown as a red bar, the box extending from the first to the third quartiles, and the whiskers extending across the range of $\eta$, omitting outliers.  The black diamonds indicate the value of $\eta$ obtained when instead of tracking each filament's displacement individually, the total summed displacement of all filaments from time point to time point is tracked.  These global measurements of displacement are more Gaussian (with $\eta \approx 0$) than the corresponding local measurements obtained from tracking filaments individually.}
	\label{FSNGP}
\end{figure}

Existing experimental studies of cytoquakes define them as large local displacements of the cytoskeleton probed using transmembrane attached microbeads or flexible micropost arrays, rather than as large changes in the cytoskeleton's total energy $U$ as done here \cite{alencar2016non,shi2019dissecting}.  To roughly compare our results to experiments, we make the corresponding local measurements of the displacements of individual filaments.   Rather than summing over all filaments, we track each filament individually and measure the set $\{\eta_f\}_{f=1}^{N_f}$ (where $N_f = 50$ is the number of filaments) of the non-Gaussian parameter $\eta_f$ corresponding to each filament $f$'s distribution of displacements from $300$ to $800$ $s$.  The calculation of filament displacements is described in SI Appendix, Filament displacements.  We find that the resulting distributions are heavy-tailed with values of the non-Gaussian parameter for most filaments in the range $\eta \sim 1-5$ (Figure \ref{FSNGP}).  This finding is in semi-quantitative agreement with \textit{in vivo} measurements on micropost arrays, whose displacements have distributions characterized by $\eta \sim 0-7$ \cite{shi2019dissecting}.  In addition, we find the distribution of $\eta_f$ itself to be heavy-tailed, also in agreement with the micropost experiments.  We next estimated the instantaneous filament speed as the filament displacement divided by $\delta t$.  We find the typical actin filament displacement speeds ($\sim 10 \ nm / s$) to be consistent in order of magnitude with separate \textit{in vitro} experiments on disordered, contractile networks which estimate this speed as $\sim 10 - 50 \ nm / s$ \cite{linsmeier2016disordered}.  These corroborations with existing measurements suggest that our simulations of a minimal cytoskeletal model system can approximately reproduce experimentally observed cytoskeletal dynamics.  We finally mention in connection to experiments that it has recently been argued that more detailed understanding of mechanical dissipation by cytoskeletal networks should help to precisely control traction-based measurements of cellular force production \cite{kurzawa2017dissipation}.  The discovery of avalanche-like dynamics in cytoskeletal networks reported in this and previous studies may help to resolve this experimental difficulty.  

The local measurements $\{\eta_f\}_{f=1}^{N_f}$, obtained by tracking each filament individually, can be compared to global measurements, obtained by summing over every filament to obtain the total displacement.  The distribution of total displacements is closer to Gaussian, characterized by $\eta \approx 0$ for most volumes tested (Figure \ref{FSNGP}).  As with the increasing Gaussianity of $\Delta U$ for large systems, this can be attributed to the central limit theorem since many filaments were summed over to determine the total displacement. We conclude that in large systems, metrics can be heavy-tailed when measured locally but Gaussian when measured globally.  This distinction between local and global measurements may be important in interpreting future studies of anomalous statistics in cytoskeletal self-organization.  

\subsection*{Normal mode decomposition probes network's mechanical state}

Having described the statistics of the increments $\Delta U$, we next aim to connect the occurrence of cytoquakes,  defined as large values of $|\Delta U_-|$, to the cytoskeletal network's mechanical stability.  To this end we implemented a method to compute the Hessian matrix $\boldsymbol{\mathcal{H}}$ of the mechanical energy function $U$.  The eigen-decomposition of $\boldsymbol{\mathcal{H}}$ is $\mathbf{\Lambda} = \{\lambda_k\}_{k=1}^{3N}$, where $3N$ is the number of mechanical degrees of freedom in the system, which comprises $N$ ``beads'' that are used to discretize the actin filaments.   $\mathbf{\Lambda}$ is related to the mechanical stability of the cytoskeletal network: the eigenvectors $ \mathbf{v}_k$ are the normal vibrational modes of the network, and the eigenvalues $\lambda_k$ indicate the stiffness ($|\lambda_k|$) and stability ($\text{sgn}(\lambda_k)$) of the corresponding mode.  Example vibrational modes are illustrated in Movies 2-5.  We draw inspiration for studying $\mathbf{\Lambda}$ in the current context from several sources: in single-molecule molecular dynamics studies, the saddle-points of $U$ (i.e. points in the landscape with some imaginary frequencies) are associated with transition states \cite{schlick2010molecular, leach2001molecular}; studies of polymer networks show that internal stresses produce non-floppy vibrational modes even below the isostatic threshold \cite{huisman2011internal}; in simulations of glass-forming liquids, the instantaneous normal mode spectrum allows inference about proximity to the glass transition and determination of incipient plastic deformation regions \cite{cho1994instantaneous, bembenek1995instantaneous, richard2021simple}; in deep learning models for predicting earthquake aftershock distributions, it was found that certain metrics also related to stability (e.g. the von-Mises criterion) are informative model inputs \cite{devries2018deep, mignan2019one}.  

\begin{figure}[ht]
	\centering
	\includegraphics[width=\linewidth]{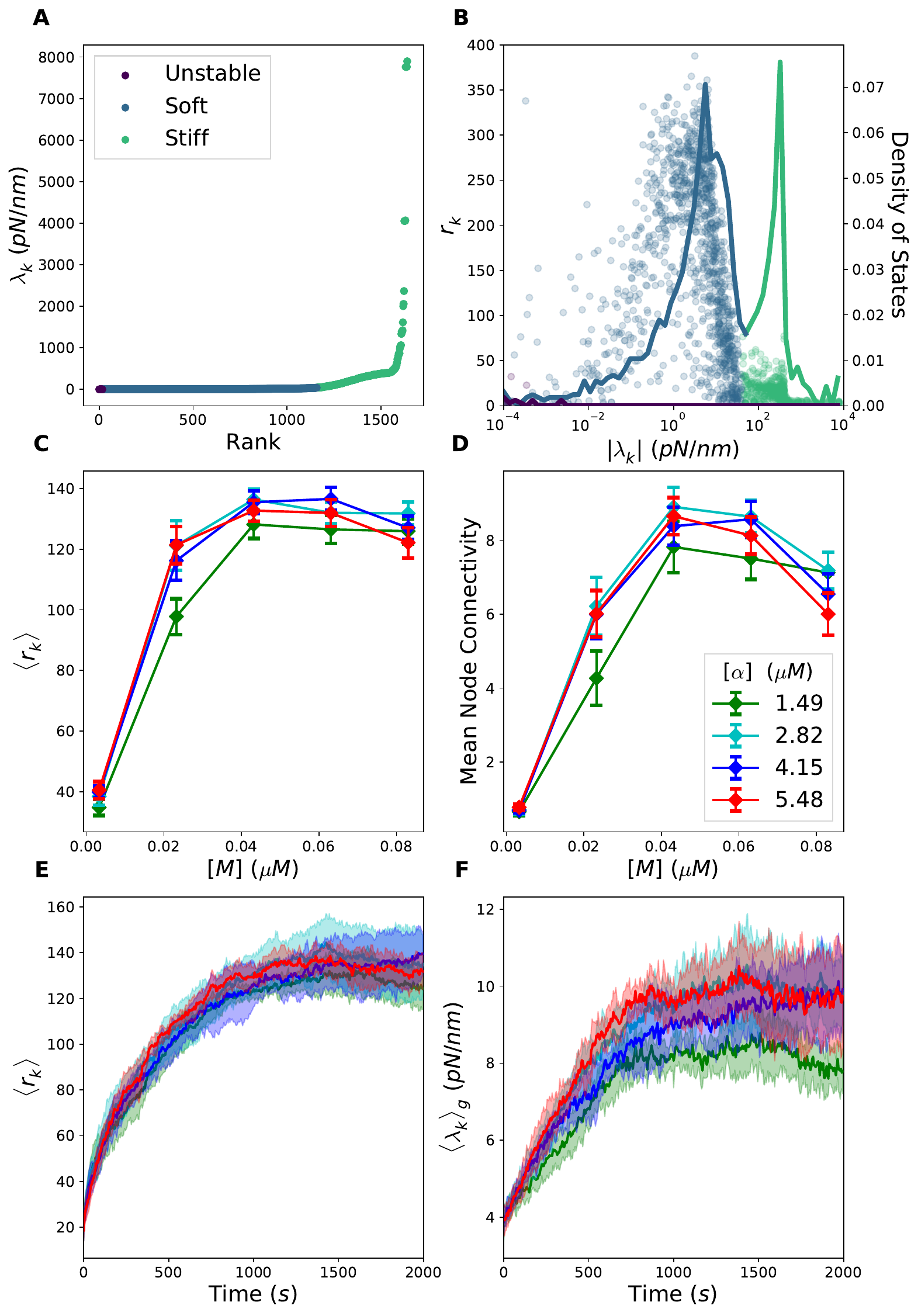}
	\caption{Metrics defined on Hessian eigen-decomposition.  \textbf{A}: Ordered eigenspectrum $\{\lambda_k\}_{k=1}^{3N}$ at a QSS time point for condition $C_{3,3}$.  \textbf{B}: Scatter plot of the pairs $|\lambda_k|, r_k$ (circles) plotted against the density of states (solid line), i.e. the proportion of eigenvalues between $\lambda$ and $\lambda + d\lambda$.  \textbf{C}: The mean value at QSS of $\langle r_k \rangle$ for the stable modes for various conditions $C_{i,j}$.  The conditions $C_{1,j}$ with low linker concentrations are not visualized as these networks did not percolate and obscure visualization for the remaining conditions.  The mean is taken over the last 500 $s$ and over different runs.  \textbf{D}: The mean value of the mean node connectivity for various conditions.  \textbf{E}: Trajectories of $\langle r_k\rangle$ of the stable modes as the network self-organizes for the conditions $C_{2,3}$, $C_{3,3}$, $C_{4,3}$, and $C_{5,3}$, where the color indicates the $\alpha$-actinin concentration as in the legend of \textbf{D}, and with the mean and standard deviation taken over the different runs.  \textbf{F}:  Similar trajectories of the geometric mean of the stable modes $\langle \lambda_k \rangle_g$.  }
	\label{HessianTrends}
\end{figure}

We briefly digress from the results on cytoquakes to describe some interesting observed trends of metrics defined on $\mathbf{\Lambda}$.  We distinguish between unstable, stable, soft, and stiff modes:  for unstable modes $\lambda_k <0$, for stable modes $\lambda_k \geq 0$, for soft modes $0 \leq \lambda_k < \lambda_T$, and for stiff modes $\lambda_k \geq \lambda_T$, where we define the threshold $\lambda_T = 40$ $pN/nm$ to discriminate between the twin peaks in the density of states (Figure \ref{HessianTrends}.B).  The set $\{\lambda_k\}_{k=1}^{3N}$ is visualized with these modes labeled in Figure \ref{HessianTrends}.A for a QSS time point of condition $C_{3,3}$.  A very small number of unstable modes persist after each minimization cycle, later iterations stopping once the maximum force on any bead in the network is below a threshold $F_T$ (here 1 $pN$).  Thus the minimized configurations are in fact saddle-points of $U$; this is expected as it is known from the theory of minimizing loss functions that the ratio of saddle-points to true local minima increases exponentially with the dimensionality of the domain \cite{dauphin2014identifying}.  We expect that in the space of all possible network topologies (i.e. patterns of cross-linkers and motors binding to filaments), the energy landscape will be rugged, leading to the well-appreciated glassy dynamics of non-equilibrium cross-linked networks \cite{wang2013microscopic,shen2004stability}.  For a fixed topology, however, which is the result of the chemical reactions occurring during step 1) of the iterative simulation cycle, the energy landscape should be smooth (i.e. not rugged) with respect to the beads' positions, with a single nearby local minimum being sought during mechanical minimization in step 3).  The residual unstable modes are therefore thought to be an unimportant artifact of thresholded stopping in the conjugate-gradient minimization routine, and not representative of some physical feature of cytoskeletal networks.  The observed quantitative dependence of the number of residual unstable modes on $F_T$ supports this conclusion and is illustrated in SI Appendix, Figure S7.

We quantify the number of degrees of freedom involved in a given normalized eigenvector $\mathbf{v}_k$ using the inverse participation ratio \cite{cho1994instantaneous}:
\begin{equation}
r_k = \left(\sum_{i=1}^{N}\sum_{\mu=1}^3 (v_{k,i\mu})^4\right)^{-1}.
\label{ipr}
\end{equation}
If the eigenmode involves only one degree of freedom, then one component of $\mathbf{v}_k$ will be one and the rest will be zero, and $r_k = 1$.  On the other hand, if the eigenmode is evenly spread over all $3N$ degrees of freedom, then each component $v_{k,i\mu} = (3N)^{-1/2}$, and $r_k = 3N$.  In Figure \ref{HessianTrends}.B we plot $r_k$ for the unstable, soft, and stiff modes along with the density of states, showing that the soft modes involve many degrees of freedom while the stiff and unstable modes are comparatively localized.  

We find that the mean value $\langle r_k\rangle$ varies non-monotonically with myosin motor concentration $[M]$ and $\alpha$-actinin concentration $[\alpha]$ (Figure \ref{HessianTrends}.C).  To understand this trend we implemented a mapping from the cytoskeletal network into a graph and measured its mean node connectivity, a purely topological measure of network percolation.  The graph is constructed to capture the cross-linker binding topology of cytoskeletal networks.  Nodes in the graph correspond to actin filaments, and weighted edges (which may be thresholded and converted to binary edges in an unweighted graph) correspond to the number of cross-linkers connecting the pair of filaments.  The mean node connectivity is defined as the average over all pairs of nodes in the unweighted graph of the number of edges necessary to remove in order to disconnect them, thus quantifying the typical number of force chains between filaments, or equivalently the extent of network percolation \cite{newman2018networks, alvarado2017force}.  Revealingly, the mean node connectivity correlates closely with $\langle r_k \rangle$ for the stable modes across the various conditions $C_{i,j}$ (Figures \ref{HessianTrends}.C and \ref{HessianTrends}.D).  We also find the number of connected components of $\boldsymbol{\mathcal{H}}$ and of the graph's adjacency matrix to match for most time points, supporting this connection between network topology and stable mode delocalization.  Intermediate concentrations of myosin motors enhance the network percolation, but as $[M]$ continues to increase the motors act to disconnect cross-linked network structures causing the mean node connectivity and $\langle r_k\rangle$ to decrease.  

We observe that as a network contracts and becomes percolated during the process of myosin-driven self-organization, the stable modes steadily delocalize ($\langle r_k\rangle$ increases) and stiffen (the geometric mean $\langle \lambda_k \rangle_g$ increases), as shown in Figures \ref{HessianTrends}.E and Figures \ref{HessianTrends}.F.  During this process we also witness a qualitative change in the level spacing statistics of the very soft and delocalized modes ($\lambda_k < 10 \  pN/nm$, $r_k > 100$) from a Poisson to a Wigner-Dyson distribution (SI Appendix, Figure S2).  This indicates that in the percolated state these vibrational modes interact and exhibit level repulsion, similar to soft particles near the jamming transition \cite{van2009jamming, silbert2009normal,  zeravcic2008localization, shi2019dissecting}.  Future studies may reveal further similarities between these systems and other marginally stable solids \cite{shen2004stability, wang2011communication}.  

\subsection*{Cytoquakes are preceded by mechanical instability and deform along soft modes}

Can the eigen-decomposition of the Hessian matrix be used to forecast cytoquake occurrence?  Intuition suggests that, by analogy with the connection between imaginary frequencies (i.e. unstable modes) and molecular transition states, the vibrational modes of the cytoskeletal network may contain information that a large structural rearrangement is poised to occur \cite{schlick2010molecular, leach2001molecular}.  To test this idea, and without detailed \textit{a priori} knowledge about which features in $\mathbf{\Lambda}$ would be informative, we implemented a machine learning model using the eigen-decomposition as the input and outputting the predicted probability of observing a large event of energy release ($\Delta U < - 100 \ k_B T$) occurring within the next $0.15$ $s$.  As detailed in SI Appendix, Machine learning model, we found that, indeed, the Hessian eigenspectrum $\mathbf{\Lambda}$ contains sufficient information to forecast cytoquake occurrence with significant accuracy compared to a random model.  We first reduced the dimensionality of $\mathbf{\Lambda}(t)$ using principal component analysis, finding that 30 dimensions sufficed to explain $> 95 \% $ of the variance across time points, and then used the reduced input in a three layer feed-forward neural network.  We validated our model using receiver operating characteristic curves, achieving an area under the curve (AUC) of $0.70$ when using data from five runs of condition $C_{3,3}$.  This improvement in prediction performance over a random model (which would have an AUC of $0.5$) implies that mechanical instability, as encoded in the Hessian eigenspectrum, precedes the occurrence of cytoquakes.

\begin{figure}[ht]
	\centering
	\includegraphics[width=1.0\linewidth]{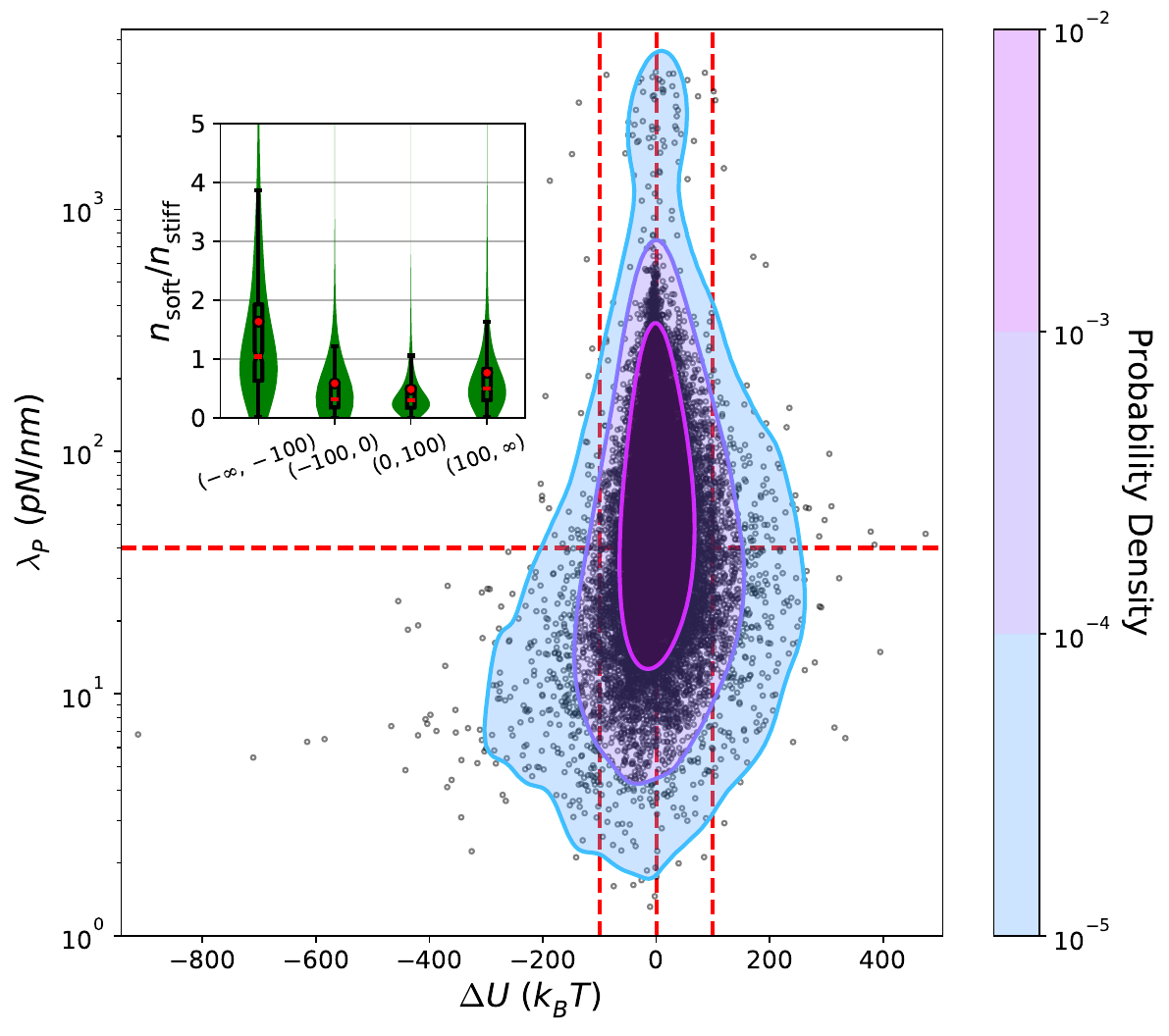}
	\caption{Scatter plot of the pairs $\Delta U,$ $\lambda_P$ measured during QSS for a run of condition $C_{3,3}$.  From these points, a Gaussian kernel density estimate of the joint PDF (treating $\lambda_P$ on a log-scale) is constructed and shown as a contour plot.  Red guidelines demarcate regions of interest.  Inset:  Combination violin and box-and-whisker plots showing the ratio $n_\text{soft} / n_\text{stiff}$ for different categories of simulation cycles, c.f. Figure \ref{ViolinPlots}.  The inset is not blocking any of the scatter plot data.   }
	\label{Projections}
\end{figure}

To further study the connection between cytoquakes and mechanical stability, we measured the projections of the network's displacements onto the vibrational normal modes $\{\mathbf{v}_k\}_{k=1}^{3N}$.  Network displacements $\mathbf{d}(t)$ were found by tracking the movement of each of the $N(t)$ beads during simulation cycles.  As a working approximation, beads that depolymerized during a cycle were assigned a displacement of $0$, and beads that newly polymerized were not assigned elements in $\mathbf{d}(t)$.  The $3N$-dimensional displacement vectors $\mathbf{d}$ were then normalized to have unit length.  We define 
\begin{equation}
d_k  =\mathbf{d} \cdot \mathbf{v}_k
\label{dk}
\end{equation}
as the projections of $\mathbf{d}$ onto the eigenmodes $\mathbf{v}_k$, which obey $\sum_{k} d^2_k = 1$ owing to the normalization of $\mathbf{d}$ and $\mathbf{v}_k$.  Thus the quantity $d^2_k$ is the weight of the displacement $\mathbf{d}$ along the $k^\text{th}$ eigenmode.   With this we define the effective stiffness
\begin{equation}
\lambda_P = \sum_{k} d^2_k \lambda_k
\label{effstiff}
\end{equation}
as the displacement-weighted average of the eigenvalues.   In Figure \ref{Projections} we display a scatter plot of the pairs $\Delta U(t),$ $\lambda_P(t)$ measured during QSS for a run of condition $C_{3,3}$, along with a kernel density estimate of their joint probability density function (PDF).  We again distinguish between soft ($0\leq \lambda_k < \lambda_T$) and stiff ($\lambda_k \geq \lambda_T$) eigenmodes, where $\lambda_T = 40 \ pN/nm$ separates the twin peaks in the density of states (see Figure \ref{HessianTrends}.B).  The structure of the joint PDF is markedly asymmetric about $\Delta U = 0$ and shows that $\lambda_P$ during cytoquake events is almost always soft, whereas for all other simulation cycles $\lambda_P$ could be soft or stiff with similar probabilities.  Because soft modes inherently involve a large number of degrees of freedom as illustrated in Figure \ref{HessianTrends}.B, we also consider 
\begin{equation}
n_k = \frac{d^2_k}{r_k}
\label{nk}
\end{equation} as the weight of the displacement along eigenmode $k$ per degree of freedom involved in the eigenmode, where  $r_k$ is the inverse participation ratio defined in Equation \ref{ipr}.  We define $n_\text{soft}$ and $n_\text{stiff}$ as the mean of $n_k$ over the soft and stiff subsets.  Values of $n_\text{soft} / n_\text{stiff}$ for different simulation cycle types are displayed in the inset of Figure \ref{Projections}, showing that $n_\text{soft} > n_\text{stiff}$ typically only during cytoquakes.  Based on this analysis, we conclude that during the large collective rearrangements corresponding to cytoquakes, cytoskeletal networks exhibit enhanced displacement along the soft vibrational modes.   We qualify these results by observing that, since cytoquakes involve particularly large network displacements, it may be inappropriate to interpret them using the local harmonic approximation to $U$ implicit in Hessian analysis \cite{richard2021simple}.  In addition, changes in network topology from linker and motor (un)binding cannot be captured using normal mode decomposition of instantaneous network configurations.  The eigenspectrum $\mathbf{\Lambda}(t)$ still informs on the stability of the energy minimized configuration before a cytoquake, but caution should be used in interpreting the cytoquake motion from $t$ to $t + \delta t$ as decomposing cleanly into non-interacting motions along the normal modes $\mathbf{v}_k$.  We leave a detailed analysis of the anharmonicity of cytoquake deformations to future work.

\section*{Discussion}

We have presented evidence supporting the following picture of active cytoskeletal network self-organization: cytoskeletal networks explore a rugged mechanical energy landscape in a stochastic process characterized by occasional, sudden jumps out of metastable configurations \cite{wang2013microscopic,shen2004stability}.  These jumps entail non-Gaussian dissipation of mechanical energy and are accomplished by an avalanche-like process of spreading destabilization, resulting in a collective structural rearrangement and a homogenization of tension.  These collective motions have large projections along the soft, delocalized vibrational modes, and, furthermore, properties of these modes can be used to predict when such relaxation events will occur.  The key finding supporting the interpretation of cytoskeletal dynamics as avalanche-like is the marked asymmetry about $0$ in the distribution of $\Delta U$ (Figures \ref{TrajCCDF}.B, \ref{FS}.B, and \ref{FS}.C).  In addition, several key quantities including filament displacements (Figures \ref{ViolinPlots}.A and Figure \ref{RankDist}), tension delocalization (Figure \ref{ViolinPlots}.D), and effective stiffness of the motion (Figure \ref{Projections}) are distributed asymmetrically about $\Delta U =0$, supporting the picture described above.

An interesting possible interpretation of the heavy tails of $|\Delta U_-|$ is that cytoskeletal networks are at a point of self-organized criticality (SOC) \cite{bak1987self,hergarten2002self,bak1989earthquakes,turcotte1997fractals,jensen1998self}.  Technical definitions of what constitutes SOC behavior are not universally agreed upon, but we may follow the definition of Ref. \citenum{hergarten2002self} which states that SOC systems must have event size distributions that tend to a power-law in the limit of an infinite system size, and a temporal signal that integrates a pink noise process, giving $\beta = 3$ for the signal.  The observed distribution of $|\Delta U_-|$ for this system size is fit by a stretched exponential function and has $\beta = 1.72$. Further, the distributions of $|\Delta U_-|$ become increasingly Gaussian for large system sizes (Figure \ref{FS}).  We thus conclude that cytoskeletal networks for these physiological conditions display non-critical dynamics, at least as measured using the global energy release $|\Delta U_-|$.  The motor walking in the system may not be sufficiently slow to yield SOC behavior, which requires a sharp separation of time scales between slow driving and fast dissipation, and the non-conservative transfer of mechanical energy between network components may also play a role \cite{jensen1998self, howard2001mechanics, floyd2019quantifying,seara2018entropy, pun2020prediction}.  We note, however, that recent studies have indicated that branched cytoskeletal networks polymerizing against a flexible membrane can produce shape fluctuations of the membrane that exhibit true SOC, leaving open the question of whether criticality is an inherent feature of cytoskeletal dynamics \cite{cardamone2011cytoskeletal,bonilla2021reproducing}. 

Instead of scale-free fluctuations, we conjecture that there exists a finite and intrinsic scale for avalanche-like releases of energy that, when summed over sufficiently large systems to obtain the global measure $|\Delta U|_-$, yields an approximately Gaussian distribution.  An important next question is then what sets this scale and how it may be measured.  We expect that the non-conservative transfer of mechanical energy through the network is one factor that attenuates the avalanches.  This non-conservation of mechanical energy arises from damping by the cytosol, accounted for in simulation through periodic minimization of the energy following stochastic chemical activity (see SI Appendix, Description of MEDYAN simulation platform).  In the lattice-based Olami-Feder-Christensen model of earthquake systems, non-conservation of energy was shown to introduce a stretched exponential cutoff to the power-law distribution of event-sizes, supporting this idea \cite{pun2020prediction,matin2020effective}.  A rough estimate of the intrinsic energy scale of avalanches can be obtained from the standard deviation of the approximately Gaussian distributions of $|\Delta U|_-$ for large systems.  However, a detailed measurement of the spatio-temporal scale will require spatially resolving the measured energy fluctuations, which was not done in this study.  In addition, in this study the temporal extent of avalanches is assumed fixed at the smallest resolved time interval $\delta t = 0.05 \ s$ (see SI Appendix, Dependence on $\delta t$ and $F_T$ for a discussion of how varying $\delta t$ affects the distribution of $|\Delta U_-|$).  Characterizing in-depth the spatio-temporal scales of avalanches is thus an important avenue for future work.

In addition to the question of what characterizes the spatio-temporal scale of cytoskeletal avalanches, several other open questions can be posed based on the results presented here.  First, we may ask about the role of force-sensitive reaction rates, including cross-linker unbinding and motor walking and unbinding, in modulating cytoquakes (see SI Appendix, Description of MEDYAN simulation platform for details of these reactions). The nonlinear coupling introduced by this force-sensitivity between the local tensions in the network and the local relaxation propensities are expected to strongly accentuate avalanche-like dynamics, but in this study we held the force-sensitive reaction rate parameters fixed at their physiological values.  Second, we may ask whether the harmonic approximation to the energy implicit in Hessian analysis is sufficient to describe the energy landscape and how it leads to avalanches.  The information on cytoquake dynamics obtained by projecting the network motion onto the Hessian eigenmodes revealed an asymmetry between energy release events and energy accumulation events (Figure \ref{Projections}), and a neural network model detected correlations between the Hessian eigenspectrum and the time-varying likelihood for a cytoquake to occur (SI Appendix, Machine learning model).  However, as cytoquakes are by definition large deformations of the network, we expect that the quadratic approximation will fail to accurately describe the energy landscape around a cytoquake event.  Higher order terms in the energy expansion or recently introduced nonlinear metrics of the local energy landscape such as the ``flatness parameter'' may be used in future computational investigations \cite{richard2021simple,feng2021inverse}.  Third, we may ask about the role of thermal noise in inducing cytoquake events.  In this study thermal noise enters in the stochastic non-equilibrium chemical dynamics which are simulated using a variant of the Gillespie algorithm over a reaction-diffusion compartment grid \cite{floyd2019quantifying}.  However, the mechanical minimization routine is deterministic given the instantaneous chemical state of the network.  Chemical reactions including motor walking and filament polymerization are expected to contribute the dominant structural fluctuations in these far-from-equilibrium networks, but the neglected diffusive motion of the filaments should also help the network escape from metastable configurations and modulate the frequency and scale of avalanches.  Elucidating whether cytoquakes can be thermally activated in this way remains another open direction for future work.

Finally, perhaps the most interesting open question regarding cytoquakes pertains to their possible physiological role in cell biology.  We proposed here that cytoquakes may be concomitant with a large structural susceptibility, by analogy with well-studied systems like the Ising model that have large susceptibilities to applied fields near their critical point \cite{zhuravlev2009molecular,binney1992theory}.  In this argument, the cytoskeleton may undergo large structural changes in response to small changes in the relevant mechanical or chemical signals, an amplification that would serve to enhance cellular sensitivity during dynamic processes such as chemotaxis.  This could also enhance mechanical adaptivity, an increasingly well-documented feature of cytoskeletal networks \cite{mueller2017load,banerjee2020actin,stern2020continual,tabatabai2021detailed}.  This connection between large cytoskeletal fluctuations and large susceptibility remains speculative at this stage, however, and would benefit from dedicated study.  Recent work has suggested that the branching agent Arp2/3, which was not included in the minimal model studied here, can enhance cytoquake sizes \cite{liman2020role}.  One can ask if by tuning the strength of this or other cytoquake-modulating factors the network is more or less responsive to external perturbation.  This perturbation could be introduced either mechanically, for example through a simulated or real force microscopy experiment, or chemically, through a variation of the chemical boundary conditions \cite{li2020tensile}.  Such studies should clarify whether large events in cytoskeletal dynamics serve a biologically useful purpose.  

\section*{Materials and Methods}
	
\subsection*{Simulation setup and conditions}

To computationally study cytoskeletal networks at high spatio-temporal resolution, we use the simulation platform MEDYAN \cite{popov2016medyan, floyd2019quantifying, chandrasekaran2019remarkable, ni2019turnover, li2020tensile}.  We provide an in-depth discussion of how MEDYAN works in SI Appendix, Description of MEDYAN simulation platform.  MEDYAN simulations combine stochastic chemical dynamics with a mechanical representation of filaments and associated proteins.  Simulations move forward in time by iterating through a cycle of four steps: 1) a short bust of stochastic chemical simulation using a variant of the Gillespie algorithm for a time $\delta t$, 2) computation of the new forces resulting from the reactions in step 1), 3) equilibration of the network via minimization of the mechanical energy, and 4) updating of force-sensitive reaction rates.  This protocol reflects an assumed separation of timescales between the slow chemical dynamics and the fast mechanical response, such that the mechanical subsystem is assumed to always remain near equilibrium and to adiabatically follow the chemical changes in the network.  As argued in Ref. \citenum{popov2016medyan}, supported using experimental evidence from Refs. \citenum{falzone2015entangled, kovacs2003functional, fujiwara2007polymerization}, this timescale separation holds for typical cytoskeletal networks which experience localized force deformations with fast relaxation times compared to the typical waiting time between myosin motor walking steps and filament growth-induced deformations.  
	
We performed MEDYAN simulations of small cytoskeletal networks consisting of 50 actin filaments in 1 $\mu m^3$ cubic boxes with varying concentrations of $\alpha$-actinin cross-linkers ($[\alpha]$) and of NMIIA minifilaments ($[M]$).  The boundaries of the box exert an exponentially repulsive force against the filaments with a short screening length of 2.7 $nm$. Five concentrations of $\alpha$-actinin (ranging from 0.17 to 5.48 $\mu M$) and five concentrations of myosin miniflaments (ranging from 0.003 to 0.08  $\mu M$) were used with a constant G-actin monomer concentration of 13.3 $\mu M$, in the regime of physiological concentrations \cite{milo2015cell}.  This led to a steady-state filament length distribution with mean 0.48 $\mu m$ and standard deviation 0.26 $\mu m$.  We label these conditions $C_{i,j}$, where $i= 1,...,5$ represents the rank of the cross-linker concentration and $j= 1,...,5$ represents the rank of the myosin motor concentration.  Five runs of each condition $C_{i,j}$ were simulated, each for $2000 \ s$.  The length of the simulation cycle $\delta t$ was chosen as $0.05 \ s$ for the results presented in this paper, although we explore dependence on this parameter in SI Appendix, Dependence on $\delta t$ and $F_T$.  
	
\subsection*{Entropy of spatial tension distribution}
The simulation volume of 1 $\mu m^3$ is discretized into $10^3$ cubic voxels, each $0.1 \ \mu m$ in linear dimension.  Let $i,j,k= 1,...,10$ index these voxels, which are an analysis tool and not related to the reaction-diffusion compartments used in MEDYAN.  After each simulation cycle, the mechanical components of the cytoskeletal network (i.e. the filament cylinders, the myosin motors, and the passive cross-linkers) are each under some compressive or tensile force $T_n$, where $n$ indexes the mechanical component.  There are other mechanical potentials involving these components, but we focus here only on the tensions $T_n$.  Each mechanical component has a center of mass $\mathbf{r}_n$, and we define the indicator function $\chi_{ijk}(\mathbf{r}_n)$ which is equal to $1$ if $\mathbf{r}_n$ is inside voxel $i,j,k$ and $0$ otherwise.  The total tension magnitude inside voxel $i,j,k$ is 
\begin{equation}
	\lvert T \rvert_{ijk} = \sum_n \lvert T_n \rvert \chi_{ijk}(\mathbf{r}_n).
	\label{tensionijk}
\end{equation}
The discrete non-negative scalar field $\lvert T \rvert_{ijk}$ is converted to a distribution $P_{ijk}$ by normalization:
	\begin{equation}
	P_{ijk} = \frac{\lvert T \rvert_{ijk}}{\sum_{ijk} \lvert T \rvert_{ijk}}.
	\label{pijk}
	\end{equation}
Finally, we introduce the discrete Shannon entropy of this distribution at time $t$ as
	\begin{equation}
	H(t) = - \sum_{ijk} P_{ijk}(t) \ln P_{ijk}(t).
	\end{equation}
The units of $H$ are nats, and large values indicate a homogeneous spatial distribution of tension magnitudes throughout the network.  Reported trends using this metric are found to be essentially independent of the discretization length.

\subsection*{Constructing the Hessian matrix}

In MEDYAN, semi-flexible filaments are represented as a connected sequence of thin cylinders whose joined endpoints (i.e. hinges) are called beads.  The set of potentials defining the mechanical energy of the filaments and associated proteins is outlined in the SI Appendix, Description of MEDYAN simulation platform.  The mechanical energy $U$ is a function of these beads' positions, and elements of the Hessian matrix are defined as 
	\begin{equation}
	\mathcal{H}_{i\mu, j \nu} = \frac{\partial^2 U}{\partial x_{i\mu} \partial x_{j\nu}} = - \frac{\partial F_{i\mu}}{\partial x_{j\nu}} = - \frac{\partial F_{j \nu}}{\partial x_{i \mu}},
	\label{eqHessian}
	\end{equation}    
where $x_{i\mu}$ is the $\mu^\text{th}$ Cartesian component of the position of the $i^\text{th}$ bead.  We have $\mu = x,y,z$ and $i = 1,...,N$ where $N$ is the number of beads in the network, so $\boldsymbol{\mathcal{H}}$ is a square symmetric $3N$-dimensional matrix.  The number of beads $N(t)$ will change as filaments (de)polymerize; in these simulations, at QSS a single filament of length $0.5 \ \mu m$ comprises $\sim$ 10 cylinders (11 beads), each $\sim$ $50 \ nm$ in length.  After each mechanical minimization, $\boldsymbol{\mathcal{H}}(t)$ is constructed by numerically computing the derivatives on the right of Equation \ref{eqHessian}.  The derivative $ \frac{\partial F_{i\mu}}{\partial x_{j\nu}}$ is found using a second-order central difference approximation by moving the $j^\text{th}$ bead in the $\pm \nu$ directions by a small amount and determining the changes in the force component $F_{i \mu}$ \cite{brooks1995harmonic}.  Due to issues of numerical accuracy, we do not assume the symmetry of the matrix $\boldsymbol{\mathcal{H}}$, but instead directly compute each component $\mathcal{H}_{i\mu, j \nu}$ and then symmetrize the result: $\frac{1}{2} (\boldsymbol{\mathcal{H}}^\intercal+ \boldsymbol{\mathcal{H}})  \rightarrow \boldsymbol{\mathcal{H}}$.

\section*{Acknowledgments}
 
We thank Qin Ni, Aravind Chandrasekaran, Michelle Girvan, Haoran Ni, Milo\v{s} Nikoli\'{c}, and Hao Wu for helpful discussions and editing of the manuscript.  This work was supported by the following grants from the National Science Foundation: COMBINE 1632976, CHE-1800418, DMR-1506969, and PHY-1427654. 

\newpage

\cleardoublepage
\onecolumngrid
\begin{center}
\Large{\textbf{Supporting Information Appendix}}
\end{center}

\section{Supplementary Figures}

\begin{figure}[H]
	\centering
	\includegraphics[width= 13 cm]{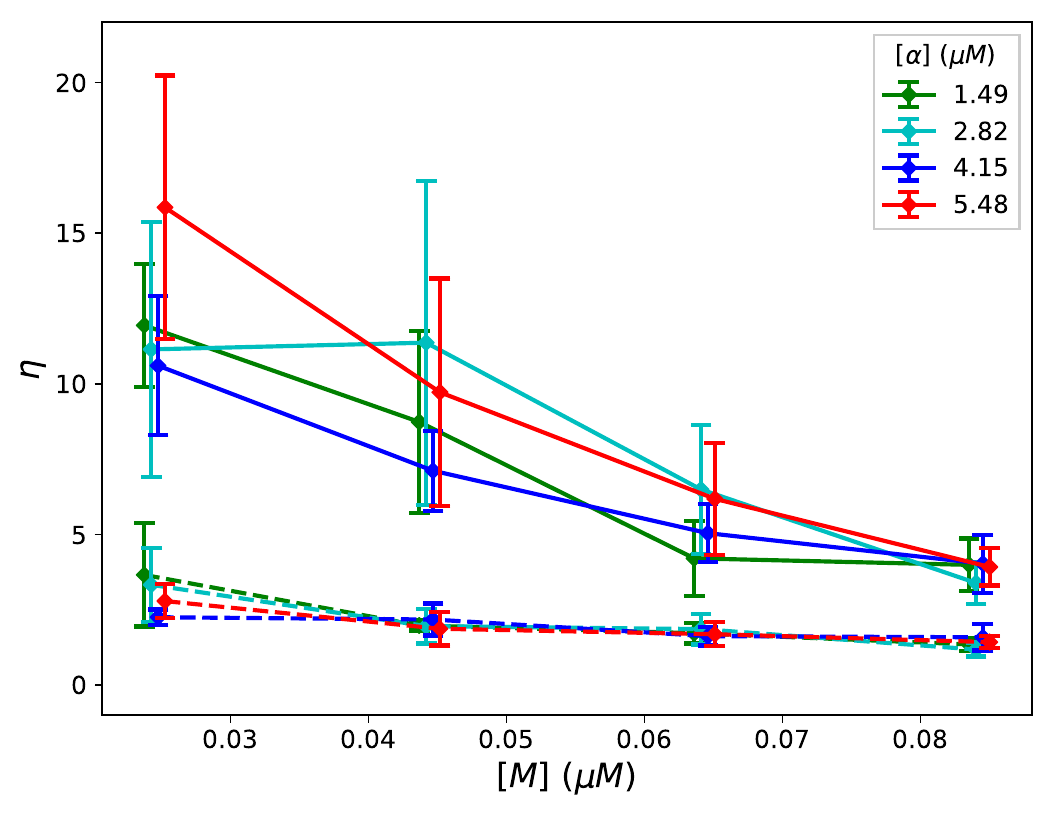}
	\caption{Plots of the non-Gaussian parameter $\eta$ for the distributions of $|\Delta U_-|$ (solid lines) and of $\Delta U_+$ (dashed lines) at QSS for various concentrations of myosin motor ($[M]$) and $\alpha$-actinin cross-linkers ($[\alpha]$).  The mean and standard deviation is shown over five runs of each condition.  A small horizontal offset is added to the points to ease visibility.}
	\label{CNGP}
\end{figure}
\clearpage
\clearpage
\begin{figure}[H]
	\centering
	\includegraphics[width= 13 cm]{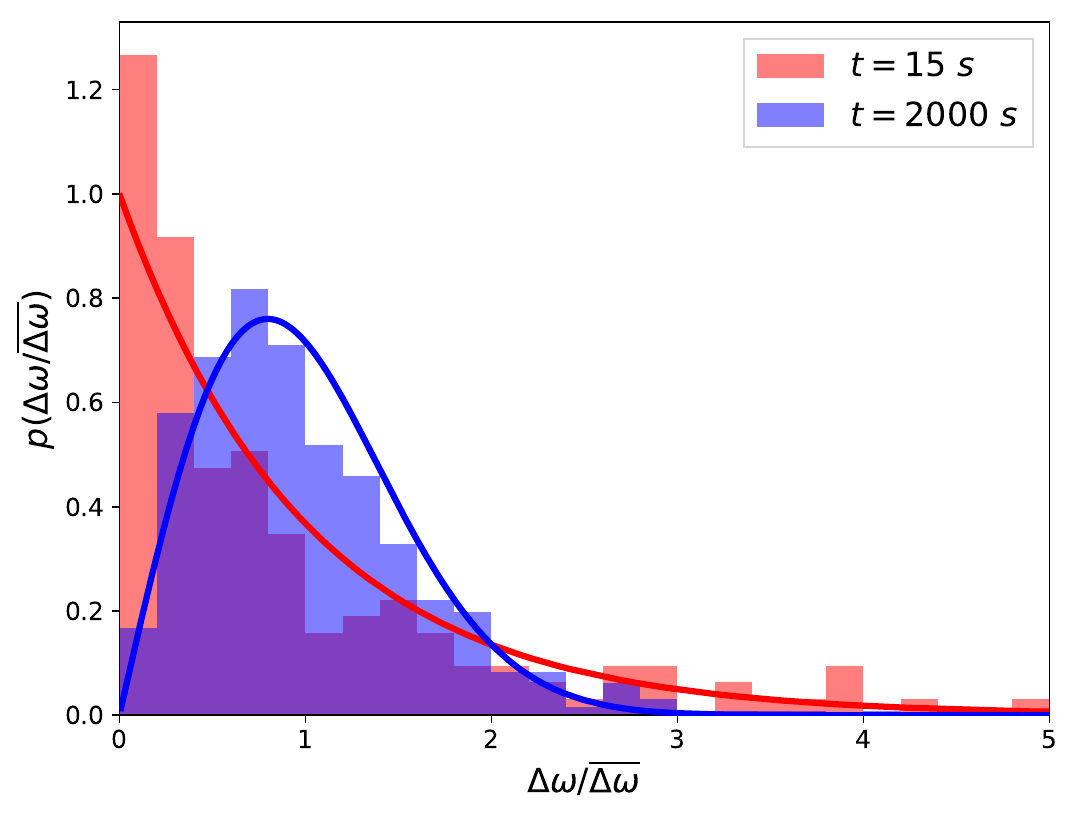}
	\caption{Histograms of the level spacings $\Delta \omega = \omega_{k+1} - \omega_k$, where $\omega_k = \sqrt{\lambda_k}$ are ordered so that $\omega_k$ increases as $k$ increases, normalized by their average $\overline{\Delta \omega}$ for the very soft ($\lambda_k < 10 \ pN/nm$) and delocalized ($r_k > 100$) vibrational modes at different times of a run of condition $C_{3,3}$.  The Poisson distribution $p(\Delta \omega / \overline{\Delta \omega}) = e^{-\Delta \omega / \overline{\Delta \omega}}$ and the Wigner-Dyson distribution $p(\Delta \omega / \overline{\Delta \omega}) = \frac{\pi}{2} (\Delta \omega / \overline{\Delta \omega}) e^{-\frac{\pi}{4} (\Delta \omega / \overline{\Delta \omega})^2 }$ are plotted as red and blue solid lines.  This transition in distributions signifies that in the percolated network at $2000 \ s$ the frequencies of these modes are no longer randomly spaced and begin to interact, exhibiting level repulsion for small $\Delta \omega / \overline{\Delta \omega}$.
	}
	\label{PoissonWigner}
\end{figure}

\clearpage
\section{Weibull plots}

\begin{figure}[H]
	\centering
	\includegraphics[width= 13 cm]{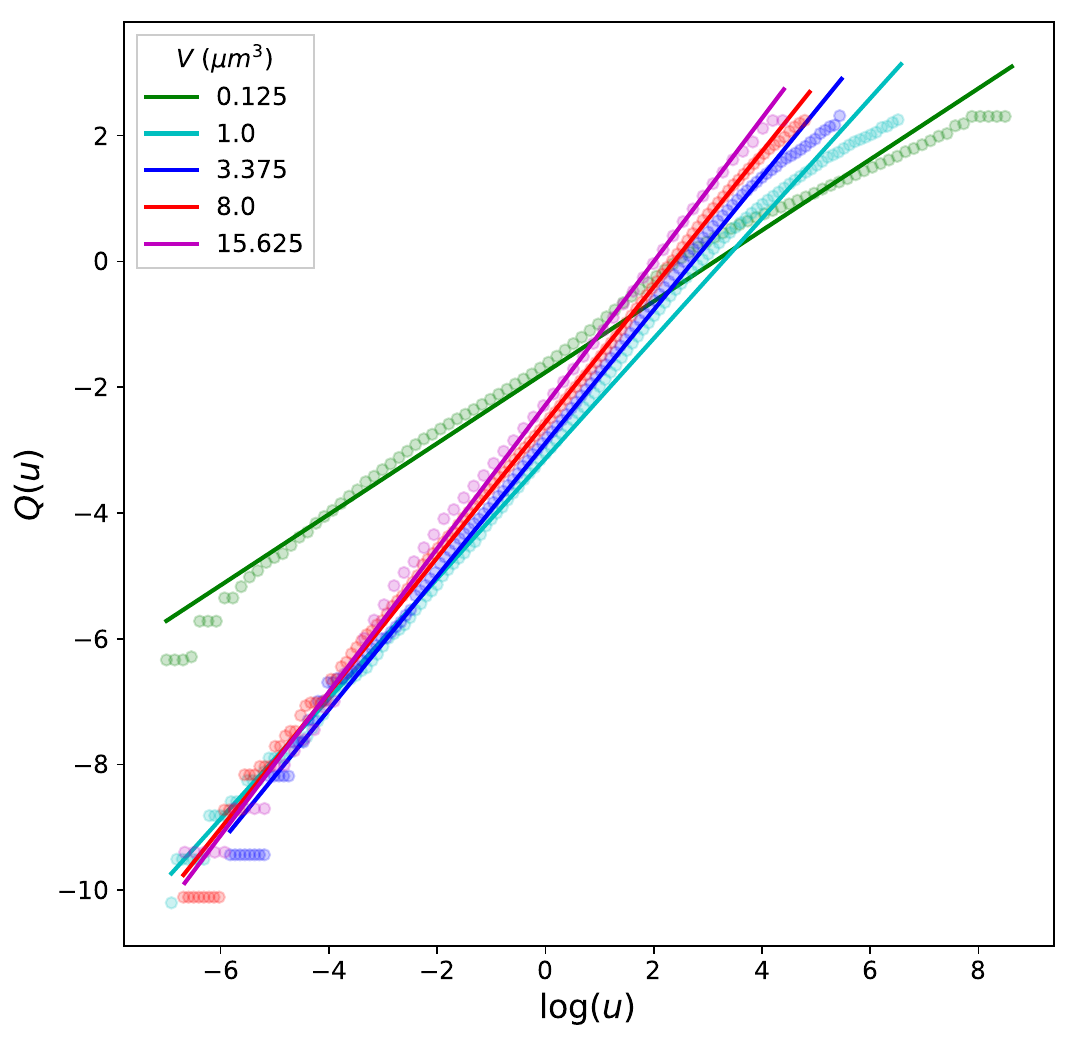}
	\caption{Plots of the function $Q(u) = \ln\left( - \ln\left(P(|\Delta U_-| \geq u) \right) \right)$ for different volumes $V$ along with a fitted line, where $P(|\Delta U_-|)$ is the observed CCDF obtained from five runs of each volume.  }
	\label{WP}
\end{figure}

\begin{figure}[H]
	\centering
	\includegraphics[width= 13 cm]{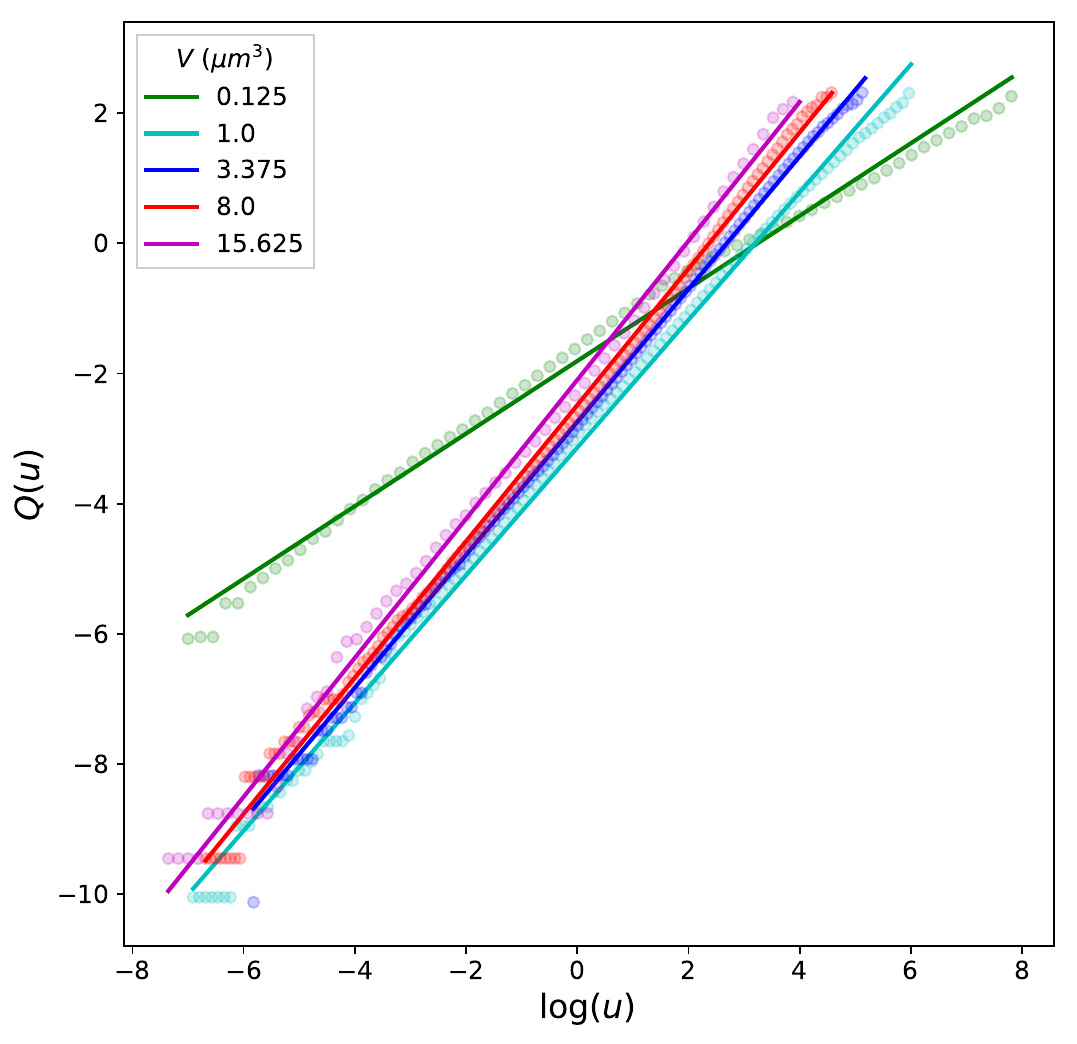}
	\caption{Plots of the function $Q(u) = \ln\left( - \ln\left(P(\Delta U_+ \geq u) \right) \right)$ for different volumes $V$ along with a fitted line, where $P(\Delta U_+)$ is the observed CCDF obtained from five runs of each volume.  }
	\label{WPPos}
\end{figure}

The degree to which the plots of $Q(u) = \ln\left( - \ln\left(P(x \geq u) \right) \right)$ against $\ln(u)$ appear to be linear serves as a check of the appropriateness of modeling $P(x)$ as a stretched exponential, or Weibull, distribution \cite{murthy2004weibull}.   See Figure \ref{WP} for $x=|\Delta U-|$ and Figure \ref{WPPos} for $x=\Delta U_+$.  On the basis of these plots we conclude that the Weibull distribution is a satisfactory choice for all values of $V$.  In the main text, the Weibull parameters $k$ and $\lambda$ were determined by fitting the stretched exponent $e^{-(x/\lambda)^k}$ to the observed CCDF $P(|\Delta U_-|)$ on a log-scale, that is, by fitting $-(x/\lambda)^k$ to $\ln\left(P(|\Delta U_-|)\right)$ using standard nonlinear fitting routines.  Treating these functions on a log-scale ensured a better fit to the distribution tails which are of most interest in the present case.

\clearpage

\clearpage

\section{Filament displacements}
\begin{figure}[H]
	\centering
	\includegraphics[width= 8 cm]{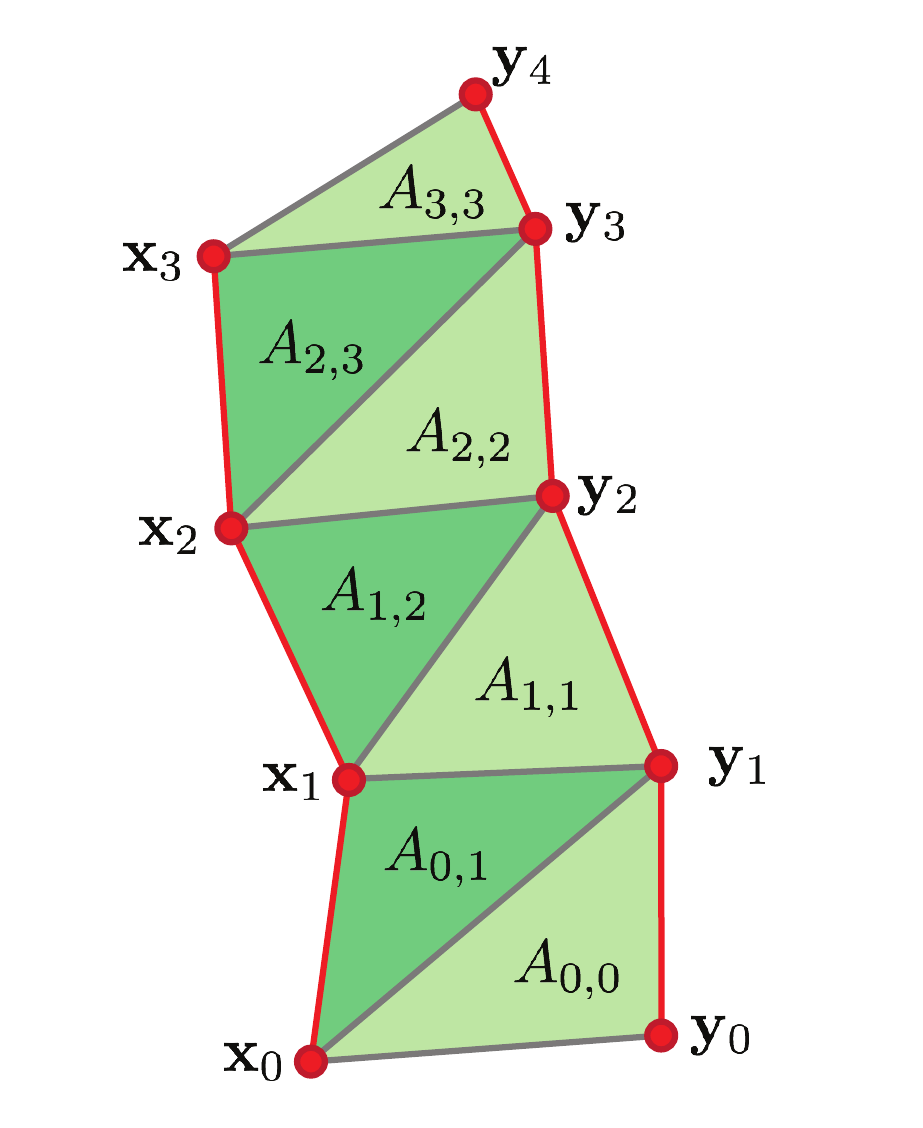}
	\caption{ Illustration of how the area between two filaments $\mathbf{x}$ and $\mathbf{y}$ is triangulated to allow calculation of the distance between them. The beads comprising the filaments are labeled $\mathbf{x}_i$, $\mathbf{y}_j$, and areas between triplets of beads are labeled $A_{i,j}$ where the lowest indices of the beads $\mathbf{x}_i$ and $\mathbf{y}_j$ in the triplet are used.  
	}
	\label{Displacement}
\end{figure}

The area between the two filaments $\mathbf{x}$ and $\mathbf{y}$ is triangulated using the beads comprising the filaments ($\{\mathbf{x}_i\}^{n_\mathbf{x}-1}_{i=0}$ and $\{\mathbf{y}_j\}^{n_\mathbf{y}-1}_{j=0}$) as vertices, where $n_\mathbf{x}$ is the number of beads in $\mathbf{x}$ and similarly for $n_\mathbf{y}$.  To compute the displacement of filament $\mathbf{x}$ during the time interval $\delta t$, we set $\mathbf{y}$ to the new configuration of $\mathbf{x}$ at the end of the interval. The triangles come in pairs for most of the filament lengths, as shown using the dark and light colors of green of Figure \ref{Displacement}.  If $n_\mathbf{x} $ and $n_\mathbf{y}$ are unequal (say $n_\mathbf{x} < n_\mathbf{y}$), extra triangles are added using the last bead in $\mathbf{x}$, $\mathbf{x}_{n_\mathbf{x}-1}$, as the only vertex in filament $\mathbf{x}$.   The sum of these triangle areas $A_\text{tot}$ is divided by the average of the two filament contour lengths $L_\mathbf{x}$ and $L_\mathbf{y}$ to give the measure of distance $d = \frac{2 A_\text{tot}}{L_\mathbf{x} + L_\mathbf{y}}$.   

\clearpage

\section{Description of MEDYAN simulation platform}
A detailed introduction to the MEDYAN (Mechanochemical Dynamics of Active Networks) model can be found in Ref. \citenum{popov2016medyan}, and additional extensions and applications of MEDYAN to study the dynamics of actomyosin networks are described in Refs. \citenum{floyd2019quantifying, chandrasekaran2019remarkable, komianos2018stochastic, ni2019turnover, li2020tensile, floyd2020gibbs,ni2021membrane,floyd2021segmental}.  Here we outline the relevant aspects of MEDYAN to facilitate understanding the results in this paper, and direct the reader to the above references for a more thorough description.  

\subsection{Simulation protocol}
A MEDYAN simulation proceeds by iterating a cycle of four steps which propagate the chemical and mechanical dynamics forward while maintaining a tight coupling between the two.  The steps are as follows:
\begin{enumerate}
	\item Evolve system using stochastic chemical simulation for a time $\delta t$.
	\item Compute the changes in the mechanical energy resulting from the reactions that occurred in step 1).
	\item Mechanically equilibrate the network in response to the new stresses from step 2).  
	\item Update the reaction rates of force-sensitive reactions based on the new tensions from step 3).  
\end{enumerate}
Further details related to these four steps are provided next.

\subsection{Chemistry}
In MEDYAN, diffusing chemical species are represented with discrete copy numbers belonging to several compartments, which form a regular grid comprising the simulation volume.  The compartment size is chosen so that it may be assumed that inside the compartments the diffusing species are well-mixed, allowing the use of mass-action kinetics to determine their instantaneous propensities to participate in chemical reactions within compartments and diffusion events between adjacent compartments.  The minimum Kuramoto length (i.e. the mean free diffusional path length of a reactive species before it participates in a chemical reaction) among the species sets this compartment size to ensure that the well-mixed assumption holds \cite{wolkenhauer2008modelling}.  The diffusing chemical species may participate in local chemical reactions according to the copy numbers of the reactants in its compartment, or else they may jump to an adjacent compartment in a diffusion event with a propensity determined by its copy number in the original compartment \cite{bernstein2005simulating}.  The algorithm for stochastically choosing which event (including local reactions or jumps between compartments) will occur next is the Next Reaction Method,  an accelerated variant of the Gillespie algorithm \cite{bernstein2005simulating, gillespie1977exact}.  These are Monte Carlo methods which randomly select both the time to any next event and which event will occur at that time in accordance with each event's instantaneous propensity.  

The user specifies the different chemical species and the reactions that they participate in.  Several types of reactions are possible.  Regular reactions involve only diffusing species (e.g. the conversion of ADP-bound to ATP-bound G-actin monomer).  Polymerization reactions result in the subtraction of a diffusing monomer from the local compartment and its conversion into a filament species, and depolymerization reactions do the opposite.  Filaments in MEDYAN's have definite spatial coordinates, rather than the compartment-level description of the diffusing species' positions.  This network of spatially resolved filaments is overlaid on the compartment grid, so that sections of filaments are able to react with diffusing species according to their local copy numbers.  In addition, filaments have mechanical properties which will be discussed in the next section.  A filament may react with a diffusing species such as a cross-linker (e.g. $\alpha$-actinin), branching (e.g. Arp2/3), or molecular motor (e.g. NMIIA).  Binding reactions involve a discrete set of binding sites along the filament, and they stochastically occur as chemical reaction events according to the number of those binding sites and the local copy number of diffusing binding molecules.  A bound molecular motor may participate in a walking reaction, which causes it to move one of its ends to an adjacent binding site, stretching the motor and generating forces.  Other reactions not used in this paper but encompassed by MEDYAN include filament nucleation, filament destruction, filament severing, and filament branching reactions.

\subsection{Mechanics}
The mechanical energy $U$ of networks in MEDYAN is a function of the positions of the filament beads and the lengths of the molecules bound to the filaments.  There are also potentials describing a branched filament's energy which are not included in this paper.  Filament beads mark the joined end points (i.e. hinges) of the cylinders comprising the filament.  Individual cylinders can stretch but not bend, but a bending energy term is included for pairs of adjacent cylinders.  The energy term for the stretching of cylinders is 
\begin{equation}
U_\text{str} = \frac{1}{2}K_\text{fil,str}(l - l_0)^2,
\label{eqstretch}
\end{equation}
where $l = \vert \vert \mathbf{r}_{i+1} - \mathbf{r}_i\vert \vert$ is the length of the cylinder whose beads are at positions $\mathbf{r}_{i+1}$ and $\mathbf{r}_i$, $l_0$ is the cylinder's equilibrium length, and $K_\text{str}$ is the spring constant of this harmonic potential.  The energy term for the bending of adjacent cylinders is 
\begin{equation}
U_\text{bend} = \epsilon_\text{bend}\left(1- \cos(\theta_{i,i+1})\right),
\label{eqbend}
\end{equation}
where $\epsilon_\text{bend}$ parameterizes the strength of the interaction and $\theta_{i,i+1}$ is the angle between the cylinders.  Molecules bound to pairs of filaments (e.g. $\alpha$-actinin and NMIIA) of stretched length $l_\text{bound}$ have a harmonic stretching energy term:
\begin{equation}
U_\text{bound,str} = \frac{1}{2} K_\text{bound,str}(l_\text{bound} - l_\text{bound}^0)^2,
\label{eqboundstretch}
\end{equation}
where the subscript ``bound" indicates that the variables and parameters are specific to the bound molecule.  An excluded volume interaction is included to prevent cylinders from overlapping.  The analytical formula for this interaction is complicated but can be expressed as a double integral over the two lengths of the participating cylinders $i$ and $j$:
\begin{equation}
U_{\text{vol},ij} = K_\text{vol} \int_0^1 \int_0^1 \frac{ds dt}{\lvert\lvert\mathbf{r}_i(s) - \mathbf{r}_j(t)\rvert\rvert^4},
\label{eqvol}
\end{equation}
where $\mathbf{r}_{i}(s) = \mathbf{r}_i + s(\mathbf{r}_{i+1} - \mathbf{r}_i)$ is the position along the $i$ cylinder, which is parameterized by a variable $s$ running from $0$ to $1$ along the cylinder's length.  These positions $\mathbf{r}_i(s)$ are also therefore functions of the cylinders' bead positions, $\mathbf{r}_i$ and $\mathbf{r}_{i+1}$.  Finally, an exponentially decaying boundary repulsion term prevents the filaments from poking outside the simulation volume:
\begin{equation}
U_\text{boundary} = \epsilon_\text{boundary} e^{-d_i/\lambda},
\label{eqboundary}
\end{equation}
where $\epsilon_\text{boundary}$ parameterizes the interaction strength, $d_i$ is the distance from the boundary to the nearest endpoint of the $i$ cylinder, and $\lambda$ parameterizes the interaction screening length.  

At the end of each chemical evolution cycle, the positions of the bound molecules and the filament beads will have changed due to the chemical reactions which occurred, displacing the system from near-equilibrium.  The positions of the filament beads are then updated in a mechanical equilibration cycle by minimizing the total mechanical energy function $U$.  This is accomplished using the conjugate-gradient minimization algorithm.  The minimization procedure ends when the maximum net force remaining in the network is below a user-specified force tolerance $F_T$, as result of which the system returns to near mechanical equilibrium.  
\subsection{Mechanochemical coupling}

An important facet of the dynamics of actomyosin networks is that the chemical reaction rates of the associated proteins depend on the forces they sustain: at high tension the myosin minifilaments will walk and unbind more slowly (stalling and catch-bond behavior) whereas the passive cross-linkers are modeled as unbinding more quickly under tension (slip-bond behavior) \cite{keller2000mechanochemistry, pereverzev2005two}.  These force-sensitive behaviors thus play the role of non-linearly coupling the mechanical state of the actomyosin network to its stochastic chemical dynamics.  

The myosin motors used in MEDYAN are modeled after non-muscle myosin IIA (NMIIA), which exists in the cell as a minifilament consisting of tens of individual myosin heads.  The chemical dynamics of the myosin minifilaments are based on the Parallel Cluster Model of Erdmann et al. \cite{erdmann2013stochastic, erdmann2016sensitivity}.  In this model, a myosin minifilament contains a number $N_\text{total}$ of individual myosin heads and has a binding rate to the actin filament pair equal to 
\begin{equation}
k_\text{fil,bind} = N_\text{total}k_\text{head,bind},
\label{eqheadbind}
\end{equation}
where $k_\text{head, bind}$ is the individual myosin head binding rate.  In MEDYAN, $N_\text{total}$ is uniformly randomly selected between a minimum and maximum number of heads each time a minifilament binds.  The bound myosin minifilament has a number of bound heads $N^0_\text{bound}$ under zero tension equal to the duty ratio $\rho$ times the total number of heads:
\begin{equation}
N^0_\text{bound} = \rho N_\text{total}.
\label{eqboundheads}
\end{equation}
The duty ratio is determined by the individual head unbinding rate:
\begin{equation}
\rho = \frac{k_\text{head,bind}}{k^0_\text{head,unbind} + k_\text{head,bind}},
\label{eqrho}
\end{equation}
where $k^0_\text{head,unbind}$ is the head unbinding rate under zero tension.  Under tension $F_\text{ext}$ the bound myosin minifilament has altered walking and unbinding rates as well as an altered number of bound heads.  The number of bound heads under tension is given by 
\begin{equation}
N_\text{bound}(F_\text{ext}) = \text{min} \left\{N_\text{total},\  N^0_\text{bound} + \beta \frac{F_\text{ext}}{N_\text{total}} \right\},
\label{eqbnoundtension}
\end{equation}    
where the parameter $\beta = 2.0$ is chosen to fit experimental data.  The myosin minifilament walking rate under zero tension is 
\begin{equation}
k^0_\text{fil,walk} = s \frac{1-\rho}{\rho}{k_\text{head,bind}},
\label{eqfilwalk}
\end{equation}
where $s$ is called the stepping fraction, defined as the ratio of the user-specified real distance between binding sites on the filament $d_\text{step}$ to the distance between binding sites on the computational cylinder representing the filament segment $d_\text{total}$: $s = \frac{d_\text{step}}{d_\text{total}}$.  Equation \ref{eqfilwalk} is based on the PCM and is explained Refs. \cite{erdmann2013stochastic,popov2016medyan}.  Under tension, the myosin minifilament walking rate is altered according to a formula of the Hill type:
\begin{equation}
k_\text{fil,walk} = \text{max} \left \{0.0, \ k^0_\text{fil,walk} \frac{F_\text{stall} - F_\text{ext}}{F_\text{stall} +  F_\text{ext} / \alpha}   \right\},
\label{eqfilwalktension}
\end{equation}
where the stall force $F_\text{stall}$ is the maximum tension a minifilament can sustain before it stops walking, and where $\alpha = 0.2$ is another parameter chosen to fit to experimental data.  The myosin minifilament will unbind from the pair of actin filaments under zero tension with a rate
\begin{equation}
k^0_\text{fil,unbind} = \frac{k_\text{head,bind}N_\text{total}}{\exp \left( \log\left( \frac{k^0_\text{head,unbind} + k_\text{head,bind}}{k^0_\text{head,unbind}} \right) N_\text{total}\right)-1}.
\label{eqfilunbind}
\end{equation}
This non-obvious expression is the inverse of the mean residence time of the minifilament as determined using the PCM.  Under tension, the myosin minifilament unbinding is modeled with Kramers-type catch-bond behavior:
\begin{equation}
k_\text{fil,unbind}(F_\text{ext}) = k^0_\text{fil,unbind} \max\left\{0.1, \  \exp \left( \frac{-F_\text{ext}}{N_\text{bound}(F_\text{ext})F_{0,\text{head}}} \right)\right\},
\label{eqfilunbindtension}
\end{equation}
where $F_{0,\text{head}}$ is the characteristic force a single myosin head catch-bond, and the minimum unbinding factor $0.1$ is a parameter to chosen to ensure the possibility to unbind under arbitrarily large tension.  We assume for myosin minifilaments that the stretching constant is given by
\begin{equation}
K_\text{bound,str} = K_\text{head,str} N_\text{bound},
\label{eqfilstretching}
\end{equation}
where $K_\text{head,str}$ is the stretching constant for an individual head; this equation assumes the bound heads share the load in parallel.  

The unbinding of passive cross-linkers (e.g. $\alpha$-actinin) are modeled as Kramers-type slip-bond:
\begin{equation}
k_\text{linker,unbind}(F_\text{ext}) = k^0_\text{linker,unbind} \exp \left( \frac{F_\text{ext}}{F_{0,\text{linker}}} \right),
\end{equation}
where $F_{0,\text{linker}}$ is the characteristic force of the cross-linker slip-bond.  

Finally, the actin filament will polymerize with a rate that exponentially decreases with the component of the sustained force along the polymerizing tip, $F_\text{ext}$.  This dependence is based on the Brownian ratchet model of Peskin et al. \cite{peskin1993cellular}:
\begin{equation}
k_\text{poly}(F_\text{ext}) = k^0_\text{poly} \exp \left( -\frac{F_\text{ext} }{F_{0,\text{poly}}} \right),
\end{equation}
where $F_{0,\text{poly}}$ is the characteristic force of the Brownian ratchet model, and $ k^0_\text{poly}$ is the zero-force polymerization rate.  

Any of the above characteristic forces $F_0$ may be converted to a corresponding characteristic distance $x_0$ via
\begin{equation}
F_0 = k_B T/ x_0,
\end{equation} 
where $k_B T$ is the thermal energy, casting expressions of the form $F_\text{ext} / F_0$ to the form $F_\text{ext} x_0 / k_B T$.
\newpage

\subsection{Parameterization}
The following table lists the parameters chosen for the simulations presented in this paper.

\begin{table}[H]
	\begin{center}
		
		\begin{tabular}{  p{3cm}  p{8cm}  p{3cm}  }
			\hline 
			\hline
			\multicolumn{1}{c}{\textbf{Parameter}} & \multicolumn{1}{c}{\textbf{Description}} & \multicolumn{1}{c}{\textbf{Value}}\\\hline
			\ & \centering{\textbf{General Simulation Parameters}} &\\
			\hline
			$k_B T$ & Thermal energy & $4.1$ $pN \cdot nm$ \\
			$L_\text{comp}$ & Cubic compartment side length &  $500$ $nm$  \\
			$N_x, \ N_y, \ N_z$ & Number of compartments in each dimension &  $2, \ 2,\ 2$  \\
			$L_\text{cyl}$ & Filament cylinder equilibrium length &  $54$ $nm$  \\
			$\delta t$ & Length of chemical evolution step &  $0.05$ $s$ \\
			$F_T$ & Force tolerance of mechanical minimization &  $1$ $pN$ \\
			\hline
			& \centering{\textbf{Mechanical Parameters}} &\\
			\hline
			$K_\text{fil,str}$ & Actin filament stretching constant &  $100$ $pN/nm$ \cite{popov2016medyan} \\
			$\epsilon_\text{bend}$ & Actin filament bending energy &  $1344$ $pN \cdot nm$ \cite{popov2016medyan, ott1993measurement} \\	
			$K_\text{vol}$ & Cylinder excluded volume constant &  $10^5$ $pN /nm^4$ \cite{popov2016medyan} \\
			$K_\text{head,str}$ & NMIIA head stretching constant &  $2.5$ $pN/nm$ \cite{vilfan2003instabilities} \\
			$K_{\alpha\text{,str}}$ & $\alpha$-actinin stretching constant &  $8$ $pN /nm$ \cite{didonna2007unfolding} \\	
			$\epsilon_\text{boundary}$ & Boundary repulsion energy &  $41$ $pN \cdot nm$ $^\textbf{a}$  \\
			$\lambda$ & Boundary repulsion screening length &  $2.7$ $nm$ $^\textbf{b}$ \\	
			\hline
			& \centering{\textbf{Mechanochemical Parameters}} &\\
			\hline
			$N_\text{NMIIA,bind}$ & Binding sites per cylinder for myosin motors & $8$ $^\textbf{c}$\\
			$N_{\alpha\text{,bind}}$ & Binding sites per cylinder for $\alpha$-actinin & $4$ $^\textbf{d}$ \\
			$d_\text{step}$ & NMIIA minifilament step size & $6.0$ $nm$ \cite{vilfan2003instabilities} \\
			$N_\text{min}$, $N_\text{max}$ & Range of number of NMIIA heads per minifilament & $15$, $25$ $^\textbf{e}$ \cite{billington2013characterization} \\
			$F_\text{stall}$ & Stall force of NMIIA minifilament & $100$ $pN$ $^\textbf{f}$ \\ 
			$F_{0, \text{head}}$ & Characteristic force of NMIIA catch-bond & $12.6$ $pN$ \cite{erdmann2013stochastic} \\ 
			$F_{0, \alpha}$ & Characteristic force of $\alpha$-actinin slip-bond & $17.2$ $pN$  \cite{ferrer2008measuring} \\ 
			$F_{0, \text{poly}}$ & Characteristic force of actin Brownian ratchet& $1.5$ $pN$  \cite{footer2007direct} \\
			$l_{M}$ & Equilibrium length of NMIIA minfilament & $175 - 225 \ nm$  \cite{popov2016medyan} \\
			$l_{\alpha}$ & Equilibrium length of $\alpha$-actinin & $30 - 40 \ nm$  \cite{popov2016medyan} \\
			\hline
			& \centering{\textbf{Chemical Parameters}} &\\
			\hline
			$k_\text{actin,diff}$ & Diffusion constant of actin monomer & $20$ $\mu M s^{-1}$  \cite{popov2016medyan} \\
			$k_{\alpha\text{,diff}}$ & Diffusion constant of $\alpha$-actinin & $2$ $\mu M s^{-1}$  \cite{popov2016medyan,hu2010mechano} \\
			$k_\text{motor,diff}$ & Diffusion constant of NMIIA minifilament & $0.2$ $\mu M s^{-1}$  \cite{popov2016medyan} \\
			$k_\text{actin,poly,+}$ & Actin plus-end polymerization & $11.6$ $\mu M s^{-1}$  \cite{fujiwara2007polymerization} \\
			$k_\text{actin,poly,-}$ & Actin minus-end polymerization & $1.3$ $\mu M s^{-1}$  \cite{fujiwara2007polymerization} \\
			$k_\text{actin,depoly,+}$ & Actin plus-end depolymerization & $1.4$ $s^{-1}$  \cite{fujiwara2007polymerization} \\
			$k_\text{actin,depoly,-}$ & Actin minus-end depolymerization & $0.8$ $s^{-1}$  \cite{fujiwara2007polymerization} \\
			$k_\text{head,bind}$ & NMIIA head binding & $0.2$ $s^{-1}$  \cite{kovacs2003functional} \\
			$k^0_\text{head,unbind}$ & NMIIA head unbinding under zero tension & $1.7$ $s^{-1}$  \cite{popov2016medyan,kovacs2003functional} \\
			$k_{\alpha \text{,bind}}$ & $\alpha$-actinin binding & $0.7$ $\mu M s^{-1}$  \cite{wachsstock1993affinity} \\
			$k^0_{\alpha \text{,unbind}}$ & $\alpha$-actinin unbinding under zero tension & $0.3$ $s^{-1}$  \cite{wachsstock1993affinity} \\
			
			\hline
			\hline
			
		\end{tabular}
		\caption{All parameters used in the simulations reported in this paper.}\label{table1}

	\end{center}
\end{table}

$^\textbf{a}$ - Chosen for the energy scale to be $10$ $k_B T$.  

$^\textbf{b}$ - Chosen as the the length of a G-actin monomer. 

$^\textbf{c}$ - Chosen to allow the spacing between binding sites to be roughly equal to its physiological value near $6$ $nm$ \cite{vilfan2003instabilities}.  

$^\textbf{d}$ - Chosen to allow the spacing between binding sites to be roughly equal to its physiological value near $30$ $nm$ \cite{meyer1990bundling}.  

$^\textbf{e}$ - Chosen to given an average $N_\text{total} = 20$ in approximate agreement with literature values \cite{billington2013characterization}.

$^\textbf{f}$ - A wide range of values are found in the literature for the stall force of the minifilament.  We take an order of magnitude estimate for this parameter based on the stall force of a single head (on the order of $10$ $pN$, estimated as $d_\text{step}K_\text{head,str}$ \cite{popov2016medyan}) times the number of bound heads in the minifilament (on the order of $10$).  This parameter choice is empirically valid as it yields observable network contraction.  

\section{Dependence on $\delta t$ and $F_T$}
\begin{figure}
	\centering
	\includegraphics[width= 13 cm]{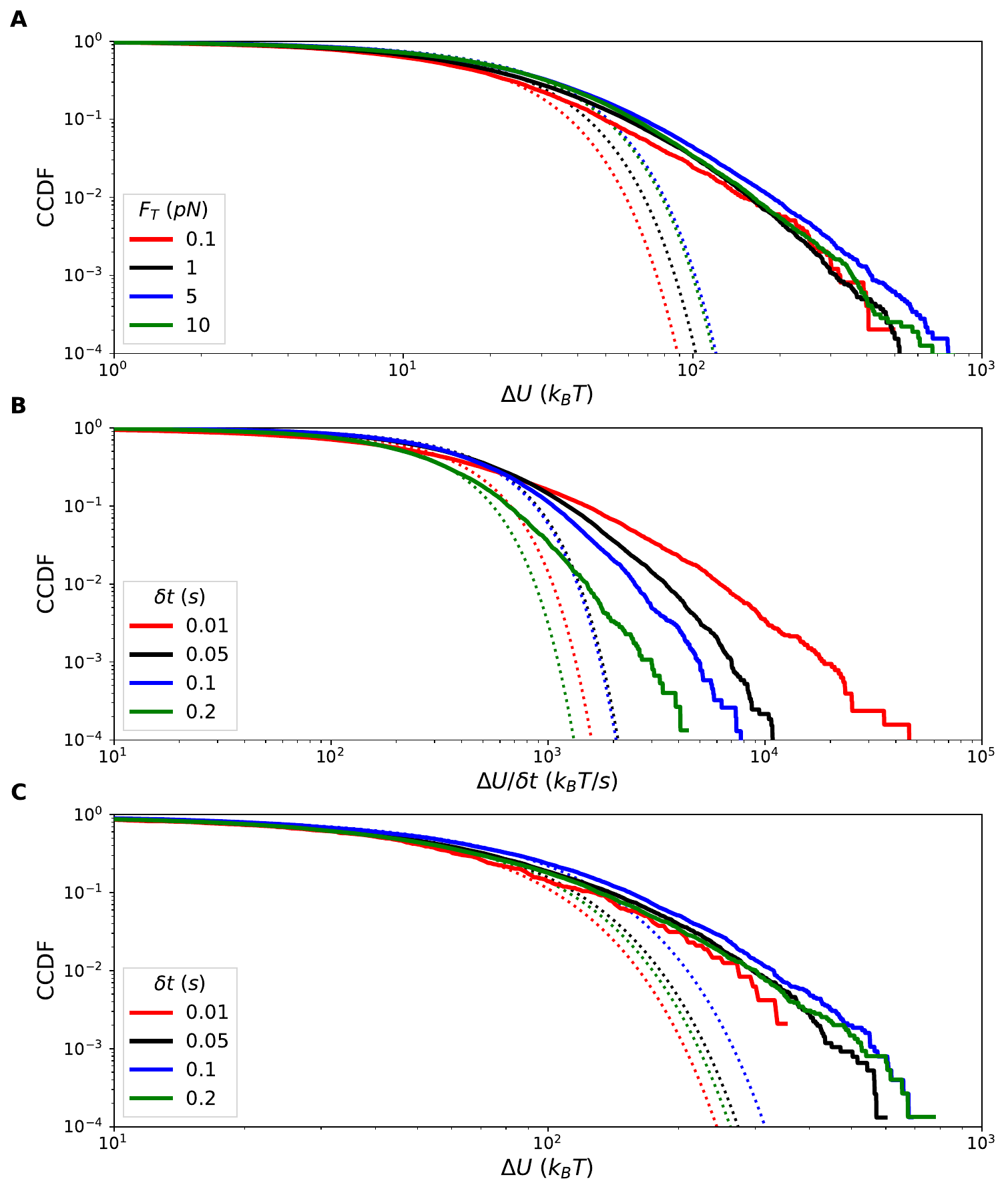}
	\caption{\textbf{A}: Complementary cumulative distribution functions of the negative increments $|\Delta U_-|$ at QSS for various choices of the force tolerance parameter $F_T$ plotted against fitted half-normal CCDFs.  For these runs condition $C_{3,3}$ is used with $\delta t = 0.05 \ s$.  \textbf{B}: CCDFs of the negative increments $|\Delta U_-|$ at QSS for various choices of the time between minimization, $\delta t$.  The energy increments are normalized by $\delta t$ for more direct comparison between these curves.  For these runs condition $C_{3,3}$ is used with $F_T = 2 \ pN$.  \textbf{B}: Complementary cumulative distribution functions of the negative increments $|\Delta U_-|$ at QSS for various choices of the time between minimization, $\delta t$.  The energy increments are normalized by $\delta t$ for more direct comparison between these curves.  For these runs condition $C_{3,3}$ is used with $F_T = 1 \ pN$. \textbf{C}:  The same data is shown as in part \textbf{B}, except here $\Delta U$ for each choice of $\delta t$ is obtained by summing consecutive energy increments over time windows equal to $0.2 \ s$.  In this way the values of $\Delta U$ for each choice of $\delta t$ correspond to the same duration of time.
	}
	\label{ControlComposite}
\end{figure}

The heavy-tailed distributions of $|\Delta U_-|$, the magnitudes of the negative energy increments which are the chief subject of this paper, may have strong dependence on certain key parameters governing the mechanical equilibration protocol.  To ensure that these distributions are not artifacts of simulation we investigate whether changing the parameters $F_T$ and $\delta t$ alters the qualitative properties of the distributions.  In Figure \ref{ControlComposite} we compare these distributions using 3 runs for each parameter choice.  Only weak dependence on $F_T$ is observed (Figure \ref{ControlComposite}.A).  We find strong dependence on $\delta t$ (Figure \ref{ControlComposite}.B), however for each parameter choice heavy tails exist and thus we may conclude that the cytoquake phenomenon is not an artifact despite their frequency and magnitude having dependence on $\delta t$.  We can ask whether the observed discrepancy between the distributions for different choices of $\delta t$ is due to a change in the underlying dynamics or due to the effect of summing over larger time intervals to obtain the quantities $\Delta U$.  We expect that by summing over larger time intervals, the heavy tails are ``averaged out,'' or coarse-grained, causing them to be increasingly Gaussian for larger $\delta t$.  We can check this by summing consecutive increments $\Delta U$ for small choices of $\delta t$ over time windows equal to the largest value of $\delta t$ tested.  When this is done (shown in Figure \ref{ControlComposite}.C), we find the distributions for all choices of $\delta t$ to approximately collapse on each other.  This suggests that coarse-graining in time indeed explains the discrepancy in the distributions of $|\Delta U|_-$ in Figure \ref{ControlComposite}.B.  Without showing the data, we find a similar picture to apply for the distributions of positive increments $\Delta U_+$, with a similar asymmetry in the non-Gaussian parameters for all choices of $\delta t$ and $F_T$ as observed for conditions used in main text, $\delta t = 0.05 \ s$ and $F_T = 1 \ pN$.  While a smaller choice for $F_T$ and $\delta t$ should correspond more closely to reality, we find that for the smallest of the tested values for these parameters the simulations did not complete in the allotted computer wall time of 2 weeks.  Thus our choices for these parameters used in this paper are chosen to be small while still allowing us to run full 2,000 $s$ simulations.   

\begin{figure}[H]
	\centering
	\includegraphics[width= 13 cm]{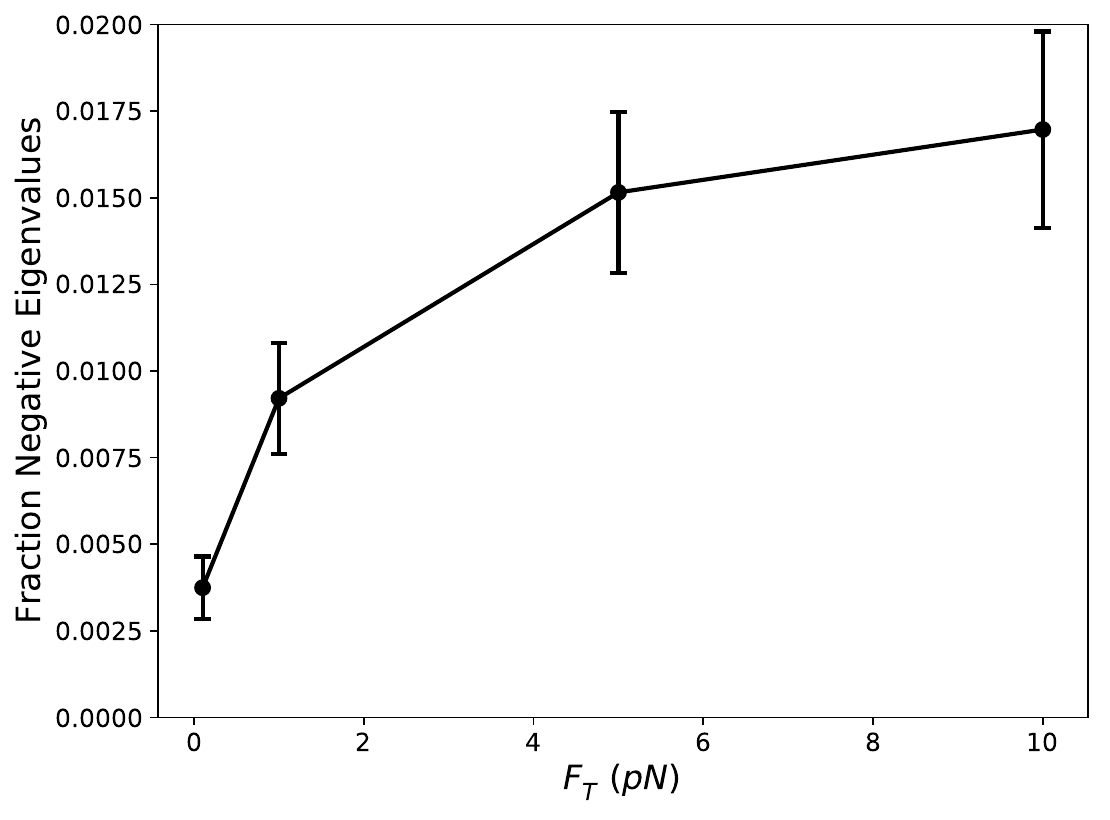}
	\caption{Scatter plot showing the fraction of negative eigenvalues remaining after mechanical minimization when different choices of the parameter $F_T$ are used.  The data is collected from QSS for 3 runs of $C_{3,3}$, with the standard deviation taken over time and over the runs.  
	}
	\label{FracNeg}
\end{figure}

\clearpage
We also investigated how the fraction of negative eigenvalues persisting after mechanical minimization depends on the force threshold $F_T$.  When minimization ceases at higher forces, more negative eigenvalues are left remaining, as expected.  This behavior is illustrated in Figure \ref{FracNeg}.

\clearpage

\section{Machine learning model}

\subsection{Cytoquake classification}
To forecast the occurrence of cytoquakes, we resorted to using a high-dimensional ML model (3 layer feed-forward neural network) after it was found that several simple features in the eigenspectrum which we believed might reflect mechanical stability (for instance the value of the smallest positive eigenvalue) did not by themselves significantly correlate with cytoquake occurrence.   We pose the forecasting of cytoquakes as a binary classification problem.  A trajectory $\Delta U(t) = U(t+\delta t)-U(t)$ at QSS (after 1,000 $s$) is converted to a binary sequence such that each $t$ for which $\Delta U(t) \leq \Delta U_T$, as well as the $t_W = 0.15$ previous seconds (i.e. 3 previous time points) are classified as cytoquakes, and the rest are not.  This $t_W$ window is chosen to help overcome the stochasticity inherent in the chemical dynamics which, along with the instantaneous mechanical stability we are using as a predictor, controls cytoquake occurrence.  We focus here on the five runs of conditions $C_{3,3}$.  $\Delta U_T = - 100 \ k_B T$ is chosen to lie well in the tail of the distribution of $|\Delta U_-|$ for this condition and therefore distinguishes rare events, as shown in Figure 1 in the main text.  With these choices, $\sim$ $10 \%$ of samples across all runs are labeled as events in the classification problem. 

\subsection{Model inputs}

\begin{figure}
	\centering
	\includegraphics[width= 13 cm]{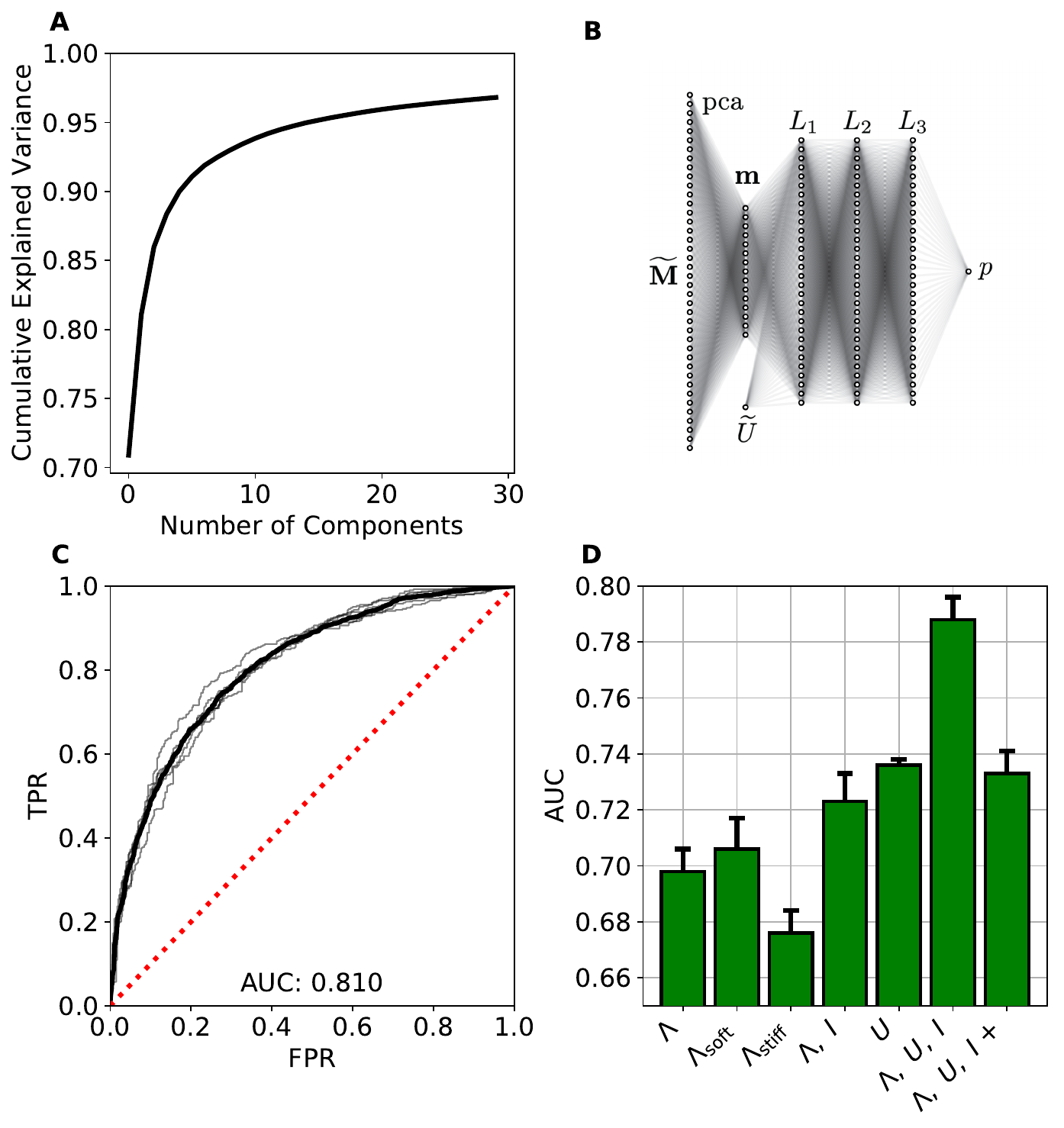}
	\caption{ \textbf{A}: Cumulative explained variance from PCA of the $\sim$ $1,600$ eigenvalues $\{\lambda_k(t)\}_{k=1}^{3N}$.  \textbf{B}: Schematic depiction of the feed-forward neural network architecture.  \textbf{C}:  ROC curves for the model using only $\{\lambda_k\}_{k=1}^{3N}$ as input and trained on a single run of condition $C_{3,3}$, with five realizations of the stochastic batch network training and their average shown.  The ROC curve of a random model is plotted as the red dotted line.  \textbf{D}:  Bar plot indicating the AUC of ROC curves using different combinations of inputs for the model trained on data collected from all runs of condition $C_{3,3}$.  From left to right, the labels indicate that the model inputs are: $\{\lambda_k\}_{k=1}^{3N}$; $\{\lambda_k | 0 \leq \lambda_k < \lambda_T\}$; $\{\lambda_k |  \lambda_T \leq \lambda_k  \}$; $\{\lambda_k\}_{k=1}^{3N}$ and $\{r_k\}_{k=1}^{3N}$; $U$, using a logistic regression model; $\{\lambda_k\}_{k=1}^{3N}$,  $\{r_k\}_{k=1}^{3N}$, and $U$; $\{\lambda_k\}_{k=1}^{3N}$,  $\{r_k\}_{k=1}^{3N}$, and $U$ with forecasting done for large positive increments $\Delta U > 100 \ k_B T$.  Error bars indicate uncertainty from five realizations of stochastic batch training.}
	\label{NNResults}
\end{figure} 

The predictors of the model capture information about the network's mechanical stability.  The ordered sets of eigenvalues $\{\lambda_k\}_{k=1}^{3N}$ at each time $t$ is padded by adding zero eigenvalues between the unstable ($\lambda_k < 0$) and stable ($\lambda_k \geq 0$) parts of the spectrum to maintain a fixed input dimension across all time points and runs.  We then collect these eigenvalues into a tuple $\mathbf{M}(t)$ such that the first element of $\mathbf{M}(t)$ is the largest negative $\lambda_k$ at time $t$ and the last element is the largest positive $\lambda_k$ at time $t$.  We optionally include the 
the inverse participation ratios $\{r_k\}_{k=1}^{3N}$ in this vector by first adding zeros in the places of the set $\{r_k\}_{k=1}^{3N}$  corresponding to where zeros were added in the set $\{\lambda_k\}_{k=1}^{3N}$, and then interleaving the $\lambda_k$ and $r_k$ in the now doubly sized tuple $\mathbf{M}(t)$, so that now for example the first two elements of $\mathbf{M}(t)$ correspond to the largest negative $\lambda_k$ and the associated $r_k$ at time $t$.  The tuples $\mathbf{M}(t)$ are then linearly rescaled, so for each element $M_i(t)$ the average over all times of a run is 0 and the variance is 1.  These rescaled tuples are labeled $\widetilde{\mathbf{M}}(t)$.  

When only the $\lambda_k$ are included then $\widetilde{\mathbf{M}}(t)$ has $\sim$ 1,600 dimensions, and with the $r_k$ are also included it has $\sim$ 3,200 dimensions.  To avoid overfitting the model, we first reduce the dimensionality of $\widetilde{\mathbf{M}}(t)$ via principal component analysis (PCA) using all QSS time points in a run.  We choose 30 dimensions as the size of the reduced tuple $\mathbf{m}(t)$ because this allows for more than $ 95 \%$ of the variance of $\widetilde{\mathbf{M}}(t)$ to be explained when just the $\lambda_k$ are included as shown in Figure \ref{NNResults}.A.  Model performance appreciably decreases when fewer than 30 dimension are used and improves only marginally if more are used.  A row of ones is added as a $31^\text{st}$ dimension to $\mathbf{m}(t)$ as a bias for the neural network.  As an additional indicator of the network's mechanical stability we also consider its mechanical energy $U$ at time $t$.   $U(t)$ is linearly rescaled to give $\widetilde{U}(t)$ so that it has zero mean and unit variance.  We then optionally augment with input tuple $\mathbf{m}(t)$ with the $\widetilde{U}(t)$ as a $32^\text{nd}$ dimension.    

\subsection{Treating multiple trials}
We can treat the data from all five runs of condition $C_{3,3}$ separately or combine all data together to train a larger model.  Model performance is generally found to be better when trained on data from a single run, however by combining data from all runs we probe more general underlying trends that are not specific to the network organization of one run.  When describing trends from varying model inputs, as in Figure \ref{NNResults}.D, we focus on results obtained by combining all runs due to their greater generality.  

For a single run there are $\sim$ 20,000 samples, giving 100,000 samples when combining all runs.  When combining runs, we first rescale and perform PCA on the predictors using only the data within a single run, and then concatenate the resulting $\mathbf{m}(t)$ with their associated labels into a larger data set.  This way the relative variation of the predictors compared to their typical values for a particular organization of the actomyosin network is retained, and the typical values of particular network organizations themselves affect the model inputs to a lesser degree.  

\subsection{Neural network architecture}
We used the Python modules scikit-learn and Keras with a Tensorflow back end to train a deep feed-forward neural network and a logistic regression model for the binary classification problem \cite{chollet2015keras, scikit-learn}.  The 31 or 32-dimensional (depending on if $\widetilde{U}(t)$ is included as a predictor) input tuple $\mathbf{m}$ is fed into three fully connected hidden layers $L_{i}$, $i = 1,2,3$, each with either 30 or 100 nodes depending on if the data consists of a single run (20,000 samples) or of all five runs (100,000 samples).  Each node in the hidden layers uses a rectified linear unit activation function.  The output of the network is two nodes using a softmax activation function whose values are $p$ and $1-p$, where $p$ is the predicted probability of a cytoquake event at that time $t$.  This architecture is schematically illustrated in  Figure \ref{NNResults}.B. The network is trained for either 400 or 200 epochs using a categorical cross-entropy loss function with Adam optimization in stochastically chosen batches of either 1,000 or 10,000 samples, depending on the whether the single or multiple run data sets, respectively, are used.  The cytoquake samples are given a higher weight ($\times 3$) than the non-cytoquake samples during training.  A L2 penalty of 0.05 is used to curb overfitting.  When using only  $\widetilde{U}(t)$ as a predictor, a logistic regression model is fit using the same sample weights.  

\subsection{Model validation}

Of all the data samples, we use $2/3$ to train the model with and validate the model on the remaining $1/3$.  We repeat these random training/testing set splits to gather statistics on model performance.  The binary classification procedure involves the probability threshold $p_T$ (such that $p>p_T$ means the model predicts a cytoquake).  Model performance is measured by varying $p_T$ from 0 to 1 and measuring the true positive rate (TPR, the proportion of actual cytoquakes correctly predicted as such) and false positive rate (FPR, the proportion of actual non-cytoquakes incorrectly predicted as cytoquakes) on the test data; the locus of these points forms the receiver operator characteristic (ROC) curve.  A random model would have FPR = TPR, so an area under the curve (AUC) of the ROC curve greater than $0.5$ indicates a good model, and a perfect model would have an AUC of 1.  One can also consider precision-recall (PR) curves, which contain points in the space of model precision (the proportion of predicted cytoquakes which were actual cytoquakes) and recall (the same as TPR).  A random model would have the same precision, equal to the proportion of actual cytoquakes in the testing data, for all values of recall as $p_T$ is varied, giving an AUC equal to that proportion.  

When the test data is unbalanced, i.e. when there are many more non-cytoquake events than cytoquake events, it has been shown that the AUC of the PR curve is a more faithful metric for model performance (since a model may score a high AUC of the ROC curve by overestimating that events are not cytoquakes) \cite{davis2006relationship, saito2015precision}.  To overcome this limitation of ROC curves, which we believe has a more intuitive interpretation that PR curves, we balance the testing data, keeping all cytoquake events and randomly keeping an equal number of non-cytoquake events.  We confirmed that trends observed in the AUC of the ROC curves as the model is varied also hold when considering the AUC of PR curves on the full test set.  

\begin{figure}[H]
	\centering
	\includegraphics[width= 13 cm]{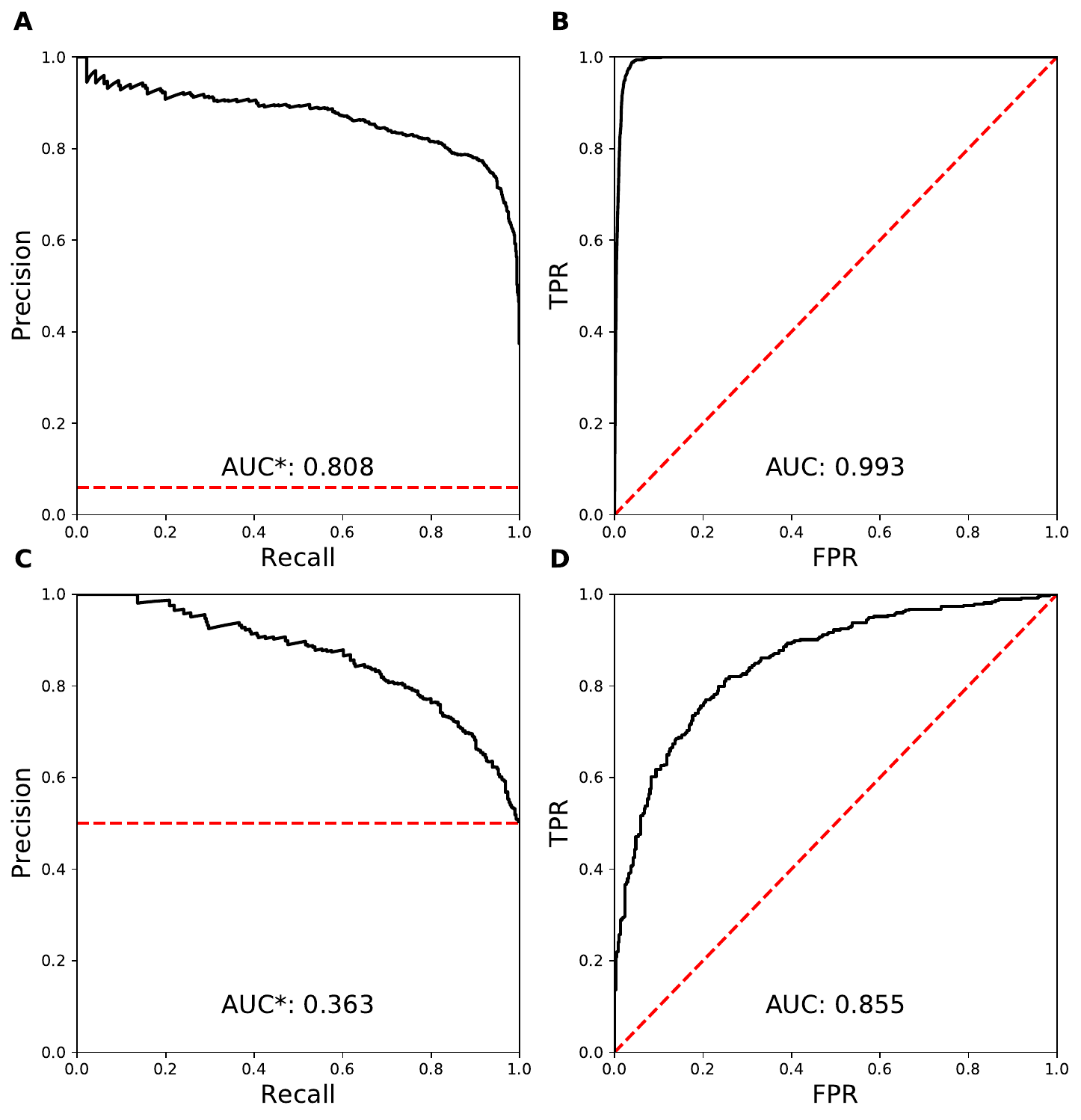}
	\caption{\textbf{A}: PR curve evaluated for a model using $\{\lambda_k\}_{k=1}^{3N}$, $\{r_k\}_{k=1}^{3N}$, and $U$ as inputs trained on data from a single run at QSS of condition $C_{3,3}$ and evaluated on the training data. The red line indicates the performance of a random model on the data set.  The asterisk on the AUC indicates that the fraction of cytoquake samples in the data set (for this run $\sim0.06$) has been subtracted from the actual AUC, to give the area between the black and red curves.   \textbf{B}: ROC curve for the same model evaluated on the training data. \textbf{C}: PR curve for the same model evaluated on the balanced testing data. \textbf{D}: ROC curve for the same model evaluated on the balanced testing data.
	}
	\label{PRROC}
\end{figure}

In Figure \ref{PRROC} we show examples of these PR and ROC curves on the training and testing data for a model trained on a single run.  The very high AUC of the PR and ROC curves evaluated on the training data indicates that the model has nearly perfected its prediction on those samples and may indicate overfitting, however this high performance generalizes nicely to the unseen testing data.  Note that the AUC of the ROC evaluated on the testing data is significantly higher than shown in  Figure \ref{NNResults}.D reflecting the generally higher performance of models trained on data from a single run compared to models trained on data from all runs.  

\begin{figure}[H]
	\centering
	\includegraphics[width= 13 cm]{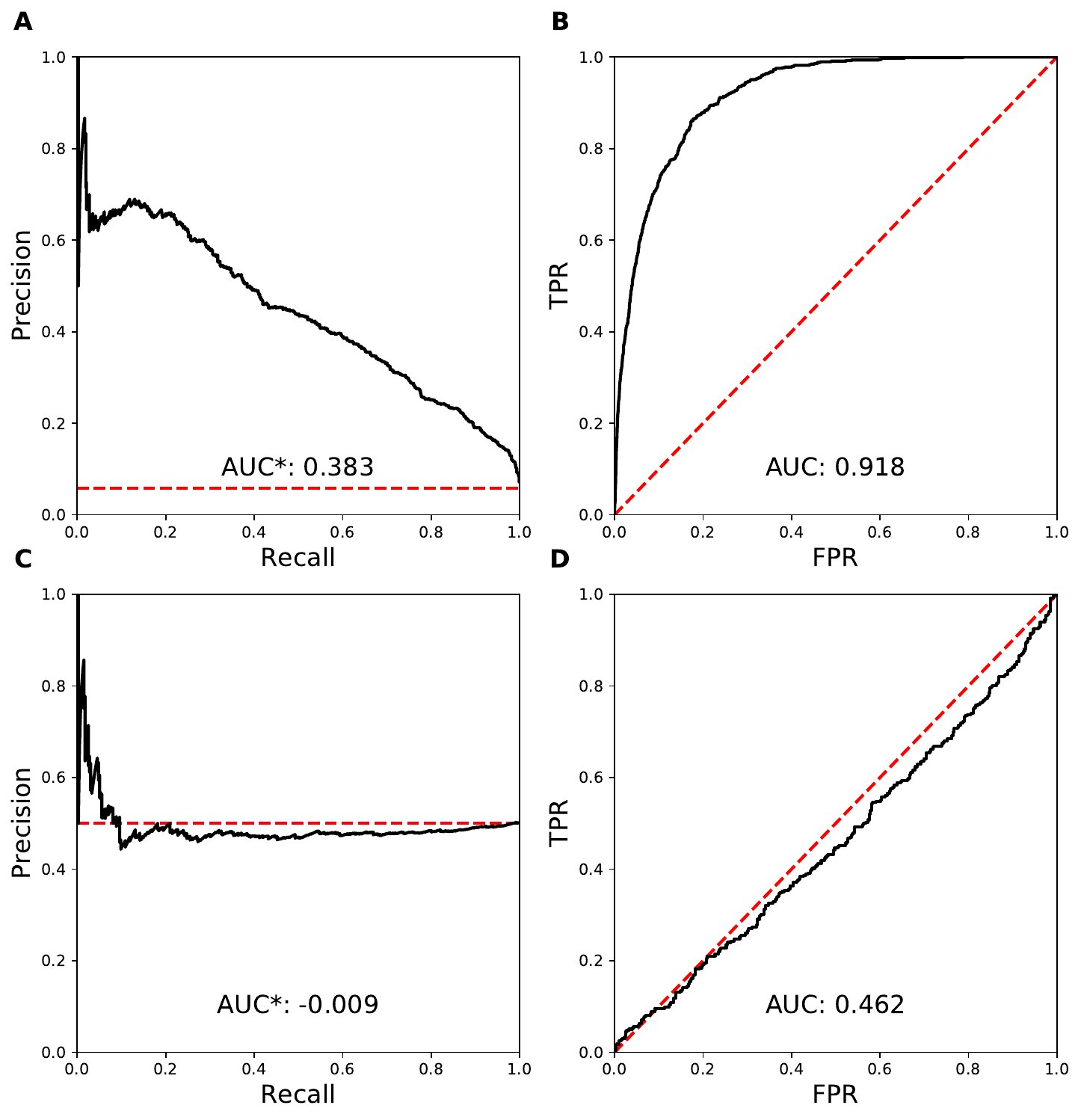}
	\caption{\textbf{A}: PR curve evaluated for a model using $\{\lambda_k\}_{k=1}^{3N}$, $\{r_k\}_{k=1}^{3N}$, and $U$ as inputs trained on data from a single run at QSS of condition $C_{3,3}$ and evaluated on the training data, when the training data labels have been randomly shuffled. The red line indicates the performance of a random model on the data set. \textbf{B}: ROC curve for the same model evaluated on the training data. \textbf{C}: PR curve for the same model evaluated on the balanced testing data. \textbf{D}: ROC curve for the same model evaluated on the balanced testing data. 
	}
	\label{ShuffledPRROC}
\end{figure}

Finally, as a sanity check, we confirmed that randomly shuffling the labels on the training set decreases performance on the training set and causes the performance on the test set to decrease to that of a random model, as shown in Figure \ref{ShuffledPRROC}.

\subsection{Varying the machine learning model inputs}

Applying the model using the Hessian eigenspectrum as the input, we obtained an AUC of 0.81 when using data from a single run of condition $C_{3,3}$ (Figure \ref{NNResults}.C) and of 0.70 when using data from five runs, i.e. from five different network realizations.  In Figure \ref{NNResults}.D, we display the effects of varying the machine learning model inputs on prediction performance, reflecting the degree to which cytoquake occurrence depends on the various inputs.  We point out that these trends from varying the model inputs are not particularly strong, contributing only marginal changes (though greater the measured uncertainty) to the model performance.  These differences are less than the difference resulting from combing all five runs in a data set rather than using one run.  We report them here mainly out of completeness, rather than in support of some strong conclusion.

Uncertainty in AUC from five repetitions of stochastic batch training is roughly $0.01$ for all reported values. Keeping only the eigenvalues of the soft modes does not harm performance (AUC 0.71), while keeping only the stiff modes does harm performance (AUC 0.68). Performance is not harmed (AUC 0.72) upon augmenting the input with the inverse participation ratios $\{r_k(t)\}_{k=1}^{3N}$.  Interestingly, we found that a logistic regression model using only the mechanical energy $U(t)$ as an input feature performs well (AUC 0.74, with a smaller uncertainty around $0.002$ for this simpler model), reminiscent of the debate concerning one neuron vs. deep learning models of earthquake aftershock prediction \cite{devries2018deep,mignan2019one}.  This logistic regression model has learned an optimal cutoff for $U$ that indicates instability and likely cytoquake occurrence.  We may seemingly conclude that the machine learning model using the Hessian eigenspectrum as an input has merely learned what the mechanical energy is, however we find that by far the best performance results from combining $\{\lambda_k(t)\}_{k=1}^{3N}$, $\{r_k(t)\}_{k=1}^{3N}$, and $U(t)$ in the ML model, reaching an AUC of 0.79 when using data from all five runs.  This suggests that the learned features of the Hessian eigenspectrum are not redundant given $U$, i.e. that their mutual information is low. Finally, we found that prediction of large positive increments ($\Delta U > 100 \ k_B T$) is also possible, with an AUC of 0.74 when combining all inputs.

\clearpage
\bibliographystyle{unsrt}

\end{document}